\documentclass{aastex}
\usepackage{emulateapj5}

\shorttitle{COLA II- Optical Spectroscopy of Southern Sample}
\shortauthors{Corbett et al.}

\begin{document}
\title{COLA II - Radio and Spectroscopic Diagnostics of Nuclear Activity in Galaxies}
\author{E. A. Corbett}
\affil{Anglo-Australian Observatory, PO Box 296, Epping 1710, NSW, Australia}
\author{L. Kewley}
\affil{Harvard-Smithsonian Center for Astrophysics, 60 Garden Street Cambridge, MA 02138}
\author{P. N. Appleton}
\affil{SIRTF Science Center, MS 220-6, California Institute of Technology, Pasadena, CA 91125}
\author{V. Charmandaris\altaffilmark{1}}
\affil{Astronomy Department, Cornell University, Ithaca NY 14853}
\author{M.A. Dopita and C.A. Heisler\altaffilmark{2}}
\affil{Research School of Astronomy and Astrophysics, Australian National University, Private Bag, Weston Creek PO, ACT 2611, Australia }
\author{R.P. Norris}
\affil{Australia Telescope National Facility, CSIRO, PO BOX 176, Epping NSW, Australia}
\author{A. Zezas}
\affil{Harvard-Smithsonian Center for Astrophysics, 60 Garden Street Cambridge, MA 02138}
\author{A. Marston}
\affil{SIRTF Science Center, MS 220-6, California Institute of Technology, Pasadena, CA 91125}

\altaffiltext{1}{Chercheur Associ\'e, Observatoire de Paris, LERMA, 61 Av. de
l'Observatoire, F-75014 Paris, France}
\altaffiltext{2}{Deceased}

\begin{abstract} 
We present optical spectroscopic observations of 93 galaxies taken from the infra-red selected COLA (Compact Objects in Low Power AGN) sample. These are all galaxies for which we have previously obtained low resolution radio
observations and high resolution ($<0.05''$) Long Baseline Array (LBA) snapshots.  The sample
spans the range of far-IR luminosities from normal galaxies to LIRGs,
and contains a significant number of galaxies involved in
galaxy-galaxy interactions. Of the galaxies observed, 78 (84\%) exhibit emission lines indicating that they are either AGN or actively forming stars (starburst galaxies). Using a newly-developed
theoretically-based optical emission-line scheme to classify the
spectra, we find 15\% of the emission-line galaxies are Seyferts, 77\% are
starbursts, and the rest are either borderline AGN/starburst or show ambiguous
characteristics. We find little evidence for an increase in
the fraction of AGN in the sample as a function of far-IR
luminosity, in contrast to previous studies, but our sample covers only a small range in infrared luminosity (10.5 $\le$ L$_{FIR}$ $\le$11.7) and thus a weak trend may be masked. Instead, as
the infrared luminosity increases so does the fraction of metal-rich
starbursts; objects which on more traditional diagnostic
diagrams would have been classified as weak LINERs (Low Ionization Narrow Emission Line Regions). As a whole the
Seyfert galaxies exhibit a small, but statistically significant, radio
excess on the radio-FIR correlation compared to the galaxies
classified as starbursts. Compact ($<$0.05'') radio cores are detected
in 55\% of the Seyfert galaxies, and these galaxies exhibit a
significantly larger radio excess than the Seyfert galaxies in which
compact cores were not detected. Our results indicate that there may
be two distinct populations of Seyferts, ``radio-excess'' Seyferts,
which exhibit extended radio structures and compact radio cores, and
``radio-quiet'' Seyferts, in which the majority of the radio emission
can be attributed to star-formation in the host galaxy. No significant
difference is seen between the IR and optical spectroscopic properties
of Seyferts with and without radio cores.

\end{abstract} 


\section{Introduction}

Many galaxies exhibit nuclear activity which cannot
be directly attributed to stars. Instead, these Active Galactic
Nuclei (AGN) are generally believed to be powered by the accretion of
matter onto a supermassive black hole
\citep[e.g.][]{Blandford92}. Recent kinematic/dynamical studies of
nearby galaxies, both with and without AGN, have revealed that many
possess a massive central dark object (see e.g. \citet{Maoz98},\citet{Ho02},
and the review paper by \citet{Kormendy01}), but it is still not known
whether all galaxies contain a central black hole. Assuming many, if not all, galaxies possess a massive nuclear black hole, an additional puzzle is that AGNs are only observed in a fraction ($\sim$ 20\%) of them.

It is often speculated that the presence of an AGN in a galaxy is in
some way related to the circumnuclear environment of the host galaxy.
However, although observations have shown that many of the host
galaxies of AGN also exhibit circumnuclear phenomena such as starburst
activity, inner rings, spirals or bars \citep[e.g.][] {Ho97b,
Martini99,Knapen00}, it is not known whether these signs of activity
are related to the mechanisms for fueling an AGN, are triggered by the
AGN, or are entirely unrelated to the presence of an AGN. Additionally
it has long been speculated that AGN-activity may be triggered and
fueled by galaxy-galaxy interactions, yet direct causal links between
the degree of interaction and AGN activity have remained elusive, at least at low luminosities \citep[e.g.][]{Bushouse86,Keel96,Wu98,Combes01}. 

Studies of the relationship between AGN and their circumnuclear environment are complicated by the need to quantify the relative strengths of the AGN and any nuclear starburst activity. These measures are often wavelength dependent and can introduce selection biases to the data. An additional problem is the nature of LINERs and their importance in a starburst/AGN regime.  

Optical line ratios have long provided a good discriminator between
AGN and starburst--dominated galaxies.  Galaxies are often
classified as AGN or starburst (HII) dominated based on their emission
line ratios, following an empirical system developed by
\citet{Baldwin81}, and refined by \citet{Veilleux87} (hereafter VO87). There are, potentially,
situations in which optical spectroscopy may fail to reveal an AGN
because of the existence of significant optical obscuration by dust in
the galaxy center. This is of particular concern for ultra-luminous
infrared galaxies (ULIRGs) and some Seyfert IIs, where large optical
and near-IR extinction is suspected
\citep[e.g.][]{Genzel98,Laurent00}. However, even in the case of very
dusty ULIRGs the optical classification of nuclear activity in all but
the most extreme cases is still reliable
\citep{Lutz98,Genzel98,Veilleux99,Hwang99,Murphy99}.  

Optical line ratios are not the only
diagnostic of AGN and starburst emission. The presence of a luminous
compact radio core in a galaxy is often taken to be a strong
indication of an AGN. Studies such as
\citet{Norris90}, \citet{Heisler98} and \citet{Kewley00} had a very high detection rate
($\sim$ 80-90\%) of compact cores ($<0.1''$) in galaxies classified
optically as AGN, but compact cores were also detected in galaxies
classified optically as starbursts albeit at a considerably lower rate
($\sim$ 5-30\%). In a much higher resolution VLBI study by \citet[][
hereafter SLL98]{Smith98a}, the detection rate of compact cores in AGN
and starburst galaxies was found to be similar, with about 50\% of the
galaxies containing cores.  While it is possible that the cores
detected in the starburst galaxies could be dust obscured AGN,
modeling by SLL98 using estimates of the star formation rate in the
host galaxies indicated that many of the lower luminosity compact
cores could be explained by emission from complexes of compact
supernovae remnants (RSNe) with a diameter of $\sim$ 10pc. Indeed
VLBI observations of Arp\,220, a source which was once believe to be
powered by a dust-enshrouded AGN, revealed that its core consists of a
complex of 12 point sources believed to be RSNe
\citep{Smith98b}. \citet{Kewley00} found that higher radio--luminosity
cores tend to be found in Seyferts rather than starburst galaxies. The
radio emission from the high luminosity cores is often more luminous
than can be accounted for by synchrotron emission from a single RSNe.

As part of the COLA project, high
resolution radio snapshot observations  of the southern sample (105) of nearby galaxies were obtained with the Australian Long Baseline Array (LBA) in order to identify compact ($<0.05''$) high brightness
temperature radio cores \citep[][ hereafter Paper I]{Corbett02}. A
study of the northern COLA galaxies is less advanced (P. Appleton et
al. in preparation), but is concentrating on the connection between
nuclear activity and galaxy-galaxy interactions. In the current paper
we present optical spectroscopic observations of 93 galaxies from the
southern sample, and use the ratios of the optical emission lines in
their nuclear (1\,kpc) regions to make spectral classifications. We
investigate whether there is any relationship between the properties
of the galaxies at radio and infrared wavelengths and their optical
classification.

The structure of this paper is as follows: Section 2 contains a brief
description of the selection criteria for the galaxy sample and the
definitions of quantities used in the paper. The optical spectroscopic
observations are presented in Section 3 together with a description of
the data reduction and calibration procedure. The optical
classifications of the galaxies are discussed in Section 4 and are
compared with those obtained for other galaxy surveys. We compare the
optical classifications of the galaxies with their infrared, far
infrared and radio properties in Section 5 and discuss the
relationship between the presence of a compact radio core and galaxy
spectral type in Section 6. Finally, in Section 7, we summarize our
conclusions. Detailed notes on individual galaxies are included in the Appendix.

\section{Sample Selection}

The COLA galaxy sample consists of an all-sky sample of galaxies in
the IRAS Point Source Catalogue (1988) which satisfy the following
criteria:

\begin{itemize}

\item Flux at 60$\mu$m, ${\rm S}_{60} > $ 4Jy. If the sources follow the radio-FIR correlation, this flux limit ensures that they have fluxes greater than 10mJy at radio wavelengths and thus the majority ($>$90\%) of the galaxies will be detected at optical
- radio wavelengths. The sample was flux limited at 60$\mu$m as AGN emission is believed to be isotropic at this wavelength \citep{Keel94};

\item 3500 $<$ helio-velocity $<$ 7000 {\rm km s}$^{-1}$ , as measured
by \citet{Strauss92};

\item Galactic latitude, $\vert b \vert$, greater than 10$^{o}$ to ensure that
the IR flux measurements were not contaminated by Galactic cirrus.
\end{itemize}

The selection criteria and potential sources of bias are discussed in detail in Paper I.

In this paper we use the standard definitions \citep[see e.g.] [] {Helou85,Sanders96} {of the infrared flux, S$_{\rm IR}$, and far-infrared flux S$_{\rm FIR}$
\begin{equation}
S_{IR} = 1.8 \times 10^{-14} (13.56 S_{12} + 5.26 S_{25} + 2.54 S_{60} + S_{100})\,\, W m^{-2}
\end{equation}
\begin{equation}
S_{FIR} = 1.8\times 10^{-14}  (2.54 S_{60} + S_{100})\,\, W m^{-2}
\end{equation}

where S$_{12}$, S$_{25}$, S$_{60}$, S$_{100}$ are the IRAS fluxes in Jy of the galaxies at 12, 25, 60 and 100$\mu$m respectively.

Throughout this paper we assume H$_{0}$=75 km s$^{-1}$ Mpc$^{-1}$ and
q$_{0}$ = 0.5 but our conclusions are independent of the exact values adopted. The absolute luminosities quoted in this paper are dependent on both H$_{0}$ and q$_{0}$ and should therefore be converted if a different cosmology is assumed. 

\section{Observations}

The optical spectroscopic observations were obtained over the period
October 1998 - February 2001 with the Dual Beam Spectrograph (DBS) on
the 2.3m telescope at the Mount Stromlo and Siding Springs
Observatory. Gratings with 1200 lines per mm were used in both the red and the
blue arms, giving a resolution of 0.57\AA\ per pixel and a wavelength
coverage of 6000-7000 \AA\ and 4600-5600\AA\ in the red and blue arms
respectively. A slit width between $1''$ and $2''$ was used during the
observations and the spatial resolution is $0.9''$ per pixel. The
seeing during these observations varied from sub-arcsecond to more
than $2''$, with $1''$ corresponding to between 0.2 and 0.4 kpc
at the redshift range of our sample. Three exposures of 300s were
initially obtained of each galaxy in our sample, but this was
increased to 600s when observing conditions
deteriorated. Spectrophotometric and smooth standards (often the same
stars) were observed during each night and used to flux calibrate and
correct for atmospheric absorption features in the spectra.

Data reduction was carried out in the standard manner using IRAF and a
pipeline reduction script. The frames were
de-biased (using both bias frames and the bias strip on the individual
observations) and flat-fielded. Galaxy spectra were extracted using an
aperture centered on the position of peak flux across the slit, and
with a width corresponding to 1\,kpc (typically 2.5-5 pixels) at the
redshift of each galaxy. If more than one peak was seen in the spatial
direction, a spectrum was extracted at the position of each
peak. Sky-subtraction was then carried out and wavelength calibration
done using an Ne-Ar arc lamp. In general an arc lamp observation was
obtained immediately after a galaxy observation, i.e.  before the
telescope was moved to its new position. In a few cases the arc
observations were obtained at the end of the night with the DBS
rotated to 0, 90, 180 and 270 degrees so as to be within 22.5 degrees
of its position during each target observation.  This condition allows
us to correct for minor wavelength shifts of the arc lines with
rotator angle caused by flexure of the spectrograph structure.
  
The observations were then corrected for atmospheric extinction and
flux calibrated using the spectrum of a spectrophotometric
standard. Finally the atmospheric absorption features in the target
spectra were removed by dividing the spectra by a template atmospheric
absorption spectrum obtained from a smooth spectrum standard observed
at a similar airmass to the target. Unfortunately, observing
conditions were rarely photometric, and on
several nights the seeing was greater than $2.5''$. As a result
the flux calibrations are not accurate to better than 30\% .

The line fluxes were measured by fitting Gaussians to the emission lines using the {\it splot}
task in IRAF. The continuum flux was subtracted using a linear fit to
small portions of the continuum either side of the emission line. The
H$\alpha$ ($\lambda$ 6562\AA) and [NII] ($\lambda\lambda$ 6583\AA, 6548\AA)
doublet were often blended and their line profiles were de-blended
by simultaneously fitting a separate Gaussian for each line. When H$\beta$ was present in absorption
as well as emission both the emission and absorption features were fitted simultaneously with Gaussians. On a few occasions, such as when the emission lines
were highly asymmetric, an adequate fit was not obtained by fitting a
single Gaussian to each emission line. In these cases, a secondary
Gaussian was introduced to each line fit.  Generally, this secondary
Gaussian had much lower flux than the principal component and a similar FWHM. For
the objects treated in this way, the flux measured from the
two-component Gaussian fit to each line was summed to give the final
flux measured for the line. One galaxy (IRAS 13035-4008) exhibited strong broad H$\beta$ and H$\alpha$ components. In this case we attempted to de-blend the
broad and narrow components with multi-Gaussian fits and used only the narrow component of the emission lines to calculate the line ratios. 

At the upper limit of the redshift range of our sample, the [SII]
doublet ($\lambda\lambda$ 6730, 6716) lies in the B-band atmospheric
absorption features (at $\sim$ 6900 \AA). It was not always possible
to achieve an adequate correction for the absorption features, and so
we do not quote the fluxes for the [SII] emission lines for these
objects.
 
The emission line fluxes measured from the fits to the spectra were
corrected for extinction by dust along the line of sight following
\citet{Veilleux87}. The Whitford reddening curve \citep[as
parameterized by][]{Miller72} was used, with the assumption that the
ratios of the line intensity at H$\alpha$ to that at H$\beta$, is 2.85
for HII-like objects and 3.1 for AGN. Each object was provisionally classified
optically as HII region-like or AGN-like (following the procedure described in Section 4.1) before
the extinction correction was applied to determine whether the AGN or HII correction was used. For objects lying close to the partition line
between the AGN and HII classes, the line fluxes were corrected once
assuming an HII-like I(H$\alpha$)/I(H$\beta$) and once assuming an AGN
I(H$\alpha$)/I(H$\beta$).  In practice the extinction correction does
not change the spectral classification and thus objects which lie on
the partition between AGN and HII are corrected for extinction using
the AGN values for the Balmer decrement. The extinction-corrected
ratios of the emission lines and the E(B-V) estimates for the COLA
galaxies are shown in Table 1.

Optical spectra were obtained for 78 of the 107 galaxies in our
initial sample. An additional 15 galaxies in our sample were observed
by \citet{Kewley01b}, and since their measurements were obtained using
an identical observing and data reduction procedure to our own, we
reproduce their results here. The remaining 14 galaxies were not observed due to a combination of poor weather and lack of telescope time. Aside from ensuring that we obtained optical spectra for those sources in which compact radio cores were detected (see Section 6) the process of deciding which galaxies to observe from the sample of 107  was essentially random. We therefore do not believe that the lack of spectra for a small number of objects will introduce a bias into our results.

We were unable to classify 15/93
galaxies either because H$\beta$ was only detected in absorption
rather than emission (6 galaxies) or because no features (either in
absorption or emission) were detected in the blue (9 galaxies),
probably due to poor signal to noise in the data. In one galaxy, IRAS
10567-4310, H$\beta$ is detected in emission but not [OIII]. An upper
limit on the [OIII]/H$\beta$ ratio has been derived for
this galaxy indicating that it is a starburst (or HII) galaxy but we do not plot it on
the optical diagnostic diagrams.  Ten galaxies
exhibit multiple ``hotspots'', and the emission line ratios for each
hotspot are listed separately in Table 1.

For these observations the slit was positioned by eye over the brightest region of the galaxy and rotated to include any obvious structures or hot spots. It is possible that inaccurate placement of the slit may have resulted in our missing the nucleus in a few cases. Given the small number of objects in which multiple regions of star formation were detected (10/93; 11\%) we would expect that inaccurate placement of the slit would be more likely to result in no emission lines being observed (and hence no classification obtained) than an incorrect classification (e.g.  a Seyfert galaxy being classified as a starburst galaxy). For this reason galaxies for which we were unable to obtain a classification are excluded from much of the statistical analysis discussed in this paper. There is a possibility, however, that small errors in slit placement may change the line ratios by a small amount. In any case, errors in positioning the slit can only reduce the degree of activity inferred from the line ratios (i.e. demote a Seyfert galaxy to a LINER or a LINER to a starburst galaxy).   

\section{Spectral Classification} 

\subsection{V95 Classification Scheme}

 The ratios of the [OIII]~$\lambda$5007/H$\beta$ lines versus the
[NII]~$\lambda$6583/H$\alpha$, [SII]~$\lambda
\lambda$6716,6730/H$\alpha$, and [OI]~$\lambda$6300/H$\alpha$ ratios
were used to classify the galaxies into starburst (HII region-like)
and AGN using the semi-empirical classification scheme developed by VO87.

The galaxies were classified for each diagnostic diagram (Fig. 1) separately
and the adopted classification is shown in Table
1. Galaxies which lie below and to the left of the partition line in each
diagram are classified as starburst (HII) galaxies.  Galaxies which
lie above and to the right of the partition line in each diagram are classified
as AGN. The AGN-like galaxies were further divided following \citet[][
hereafter V95]{Veilleux95} into ``classic Seyferts'' which exhibit
high excitation with [OIII]5007/H$\beta$ $\ge$3 and LINERS which show
lower excitation, i.e. [OIII]5007/H$\beta$ $<$3
\citep[e.g.][]{Heckman80}.  For the remainder of this paper we will
use the term V95 to refer to the VO87 classification system with the
additional LINER class. In Figure 1, the V95 partition line is shown
as a thick dotted line with thin dotted lines at $\pm$ 0.05 dex. A
thin dotted line is also shown at [OIII]5007/H$\beta$ = 3.

AGN with broad Balmer emission lines (i.e. $>$ 2000 kms$^{-1}$) were
classified as Seyfert 1 galaxies. Galaxies which lie within 0.05 dex
of the partition line (shown as dotted lines on the diagnostic
diagrams; Fig. 1) are classified as borderline or transition
objects. The final classification of each galaxy is based on the class
given for each diagnostic diagram. Galaxies classified as borderline
in at least two diagrams or borderline in one and AGN in another have
a final classification of borderline (B). A galaxy classified as HII
in two diagrams and borderline in one would have a final
classification of HII. Finally, those galaxies which are classified as
HII galaxies in one diagram and AGN in another diagram are
classified as ambiguous.

For galaxies with multiple ``hotspots'', it was necessary to define
one overall classification in order to simplify comparison with other
surveys and our lower resolution multi-wavelength observations of the
galaxies. We have therefore defined the aperture (or ``hotspot'')
closest to the center of the galaxy as the ``nuclear'' spectrum and it is
this classification which is used for the galaxy in the analysis
described in Sections 5 and 6. On one occasion it was impossible to determine which hotspot was the more central and the hotspot with the brightest emission lines was designated the ``nuclear'' spectrum.  The additional ``hotspots'' detected
represent secondary emission line regions (generally fainter than those of the ``nuclear'' hotspot), possibly due
to a galaxy merger, and may even be additional nuclei. It can be seen
from Table 2 that the majority (10/13) of these additional emission
line regions are classified as starbursts, with the remaining three classified as ambiguous (2) or borderline starburst (1). 

According to the V95 classification scheme, we find that 12 of the
galaxies in our sample are Seyferts and 45 are dominated by starburst
emission (Table 2). Excluding the 15 galaxies for which no spectral
classification was obtained, this amounts to 15.4\% Seyferts and 57.1\%
starburst with the remainder having borderline (10.2\%), LINER (7.7\%) or ambiguous (9.0\%) classifications, in good agreement with other similar studies. For example, from observations of 182
luminous infrared galaxies (L$_{\rm IR} >$3$\times 10^{10}$ L$\sun$)
V95 classified 59\% as starbursts and 14\% as Seyferts. Similarly, Kewley et al. (2001a; hereafter K01), using the same system as V95, classified 58\% of galaxies as starbursts and 18\% as Seyferts from a
sample of 225 galaxies selected on the basis of their IR colour (8
$>S_{60}/S_{25}>$0.5 and 2 $>S_{60}/S_{100}>$ 0.5) and flux (
S$_{60}>$2.5Jy). A series of papers by \citet{Ho97a} (hereafter H97),
investigates the prevalence of low luminosity AGN (so called ``dwarf''
Seyfert galaxies) from spectroscopic observations of 486 nearby
($z\sim$0) galaxies. Of the galaxies with emission line nuclei they
classify $\sim$ 12\% as Seyferts and $\sim$ 50\% as starburst
galaxies.

\subsection{K01 Classification Scheme}

Recently, K01 used detailed stellar population synthesis and
photoionization models to develop an alternative scheme for the
classification of starburst and AGN.  This scheme is based on the
standard emission line diagnostic diagrams and new theoretical
classification lines. K01 demonstrated that their new
theoretical scheme reduces the number of ambiguous classifications
when compared to the V95 scheme, concluding that the position of the
theoretical lines are more consistent from diagram to diagram than the
semi-empirical V95 method. We also used this scheme to classify the
COLA galaxies. The theoretical lines are shown as thick solid lines in
Figure 1, with thinner solid lines at $\pm$ 0.5dex.  As before,
galaxies are classified as starbursts if they lie below and to the
left of the theoretical line, and as AGN if they lie above and to the
right of the theoretical line. As with the V95 system, galaxies which lie within 0.5 dex of
the line are classed as borderline (B).
Galaxies which fall within the AGN region on one or two diagrams and
the starburst region on the remaining diagram(s) have been given
``ambiguous'' classifications.
 
Using the K01 diagnostic diagrams results in a higher fraction of the
galaxies being classified as starburst galaxies (76.9\%) than from the
V95 diagrams (Table 2). This is at the expense of the LINER,
borderline and ambiguous classes which fall to 0\%, 2.6\% and 5.1\% of
the galaxy classifications respectively.  All the galaxies classified
as AGN were also classified as Seyferts under the V95 system and will
therefore be referred to as Seyferts for the remainder of this
paper. The fact that the number of ambiguous classifications as a
whole is lower for the K01 scheme (4/93; 5\%) than the V95 scheme (7/93; 9\%)
supports the conclusion by K01 that their theoretical scheme produces
more consistent classifications from diagram to diagram. It is also
interesting to note that all six galaxies classified as LINERs under
the VO87 system are reclassified under the K01 system (3 as starburst
galaxies, 2 as ambiguous and 1 as borderline). A more detailed
investigation of the galaxies which change classification under the
different schemes is given in an companion paper (P. Appleton et al., in
preparation) but one possibility is that the original V95 "LINER"
classification actually encompasses AGN/starburst composites and metal-rich starbursts, the latter resulting in low [OIII]/H$\beta$ ratios
but high [NII]/H$\alpha$ ratios.

\subsection{AGN Contribution}

The theoretical models developed by K01 have been used recently to
develop mixing grids which allow the contribution of an AGN to the
optical emission-line ratios to be found \citep[][hereafter K02]{Kewley02}.  
Since there exists a range of possible starburst-AGN
ratios due to the various combinations of ionization parameter and 
metallicity possible, K02 defined a maximum and a minimum mixing grid
for each diagnostic diagram, which give the largest possible range in
the AGN-related contribution to the optical emission-line ratios.  The
AGN contribution was calculated for each diagnostic diagram using the
average result from the maximum and minimum mixing grids.  The final
AGN contribution used is an average from the three diagnostic
diagrams, and is given in Table 3.  The lower and upper limits on the
AGN contribution for each galaxy are given by the lowest and highest
contributions found out of all three diagrams.

It is important to note that this AGN contribution is an estimate of
the amount of AGN contribution to the optical emission line ratios
{\it only} (eg. [OIII]/H$\beta$, [NII]/H$\alpha$, [SII]/H$\alpha$, and
[OI]/H$\beta$), not an estimate of the AGN contribution to the
continuum, luminosity, or to the emission-lines by themselves.  Nevertheless,
the AGN contribution can serve as a useful indicator of the relative 
amount of star formation and AGN activity responsible for the 
radiation which ionizes the narrow emission-line gas.  The 12 galaxies 
classified as Seyferts using both the V95 and K01 schemes are marked with 
an astrix (*) in Table 3.  It is interesting to note that out of these 12 Seyfert
galaxies, 4 have optical AGN contributions under 50\%, and one galaxy
(IRAS 03022-1232) has an AGN contribution of only 25\% to the total emission.
The fact that an AGN needs to contribute only a small proportion of the emission line ratios for the galaxy to be classified as an AGN was also noted by K01, and is most likely due to the log-log nature 
of the diagnostic diagrams, making them particularly sensitive to the 
presence of an AGN, even when dominated by star formation.  
Massive star formation is generally expected near the nuclei
of infrared luminous galaxies since interactions with companions,
formation of bars, and other instabilities can efficiently drive large
amounts of gas to the center
\citep[see][]{Bryant99,Combes01}. 

Compact nuclear starbursts and AGN in IR galaxies may be enshrouded by large 
quantities of dust \citep{Clavel00}.  In these cases, the emission is 
optically thick in the optical and near-IR, and the spectral energy 
distribution peaks in the far-IR (40--100$\mu$m). As a result, even if 
the presence of an AGN is revealed in the optical, it is difficult to 
quantify the effect of dust absorption and the reprocessing of its 
energy to longer wavelengths. To measure the AGN contribution to 
the bolometric luminosity of active galaxies better, diagnostics in mid-IR wavelengths, 
where we are less affected by extinction, have been developed \citep[see][
and references therein]{Lutz98, Genzel98, Laurent00,Charmandaris02a}. Recent
results using such IR diagnostics suggest that most ultraluminous infrared galaxies 
dominated by star formation in the optical are also dominated by their star formation 
in the infrared \citep{Genzel98, Lutz99}. 
However, while observations in the mid-IR can probe the dusty cores of IR
bright galaxies much better than visible observations, the cores can still be
optically thick in the mid-infrared. Furthermore, only a small fraction
($\sim$3--5\%) of the total luminosity of the most luminous infrared
galaxies emerges in the mid-IR \citep{Charmandaris02b}\footnote{For
normal late type galaxies it has been shown that $\sim$15\% of the
luminosity is emitted between 5--20$\mu$m \citep{Dale01}}.  Ideally
one would want to directly probe the physical properties of the
nucleus in wavelengths where most of the energy of those galaxies
appears, but unfortunately the spatial resolution at far-IR is rather
poor.

\section{Infrared and Radio Properties}

\subsection{Infrared Properties}

The IR fluxes used to select the original sample \citep[see
also][]{Corbett02} were taken from the {\it IRAS Point Source
Catalogue} \citep{Beichman88}. However, these fluxes are believed
to be less accurate than those given in the {\it Faint Source Catalogue
(FSC)} \citep{Moshir90}.  These new FSC fluxes are used for the
analysis of the COLA galaxies, and are shown in Table 4.

It has long been proposed that the spectral class is linked with
infrared luminosity (L$_{\rm IR}$).  In general, AGN are found in
galaxies which exhibit higher L$_{\rm IR}$ than starburst galaxies
selected from the same volume-limited sample (e.g. Sanders88; V95).  To investigate such a dependency in the COLA sample, we split our sample into a
number of luminosity ranges (shown in Fig. 2a; Table 5). Using the K01
classification scheme, we find that the proportion of starburst and
Seyfert galaxies in each luminosity bin remains roughly constant with
luminosity. In fact the Student t-Test reveals no statistical difference between the infrared luminosity of the Seyfert galaxies and the starburst galaxies. The number of galaxies with an ambiguous classification increases from $\sim$ 0\% (0/22) with L$_{\rm IR}<$10$^{10.75}$ to $\sim$18\% (2/11) with  L$_{\rm IR}>$ 10$^{11.25}$ but the small number of ambiguous galaxies in our sample (5/93) means that this trend is probably not significant. 

\citet[][ hereafter V99]{Veilleux99} conducted an optical
spectroscopic study of 108 ultraluminous infrared galaxies (L$_{\rm
IR}> 10^{12} L\sun$) and combined it with V95 (L$_{\rm
IR}>$3$\times10^{10}$ L$\sun$). They found that the proportion of
galaxies classified as AGN (Seyferts and LINERS) increased with IR
luminosity, contrary to our findings.

One possible explanation for this discrepancy is that the V99
sample extends into the ULIRG regime (L$_{\rm IR}> 10^{12}$) whereas our sample contains only 11 objects with L$_{\rm IR}> 10^{11.25}$. Our sample may simply not cover a large enough range in luminosity to identify a weak trend or, alternatively, the onset of a larger fraction of AGN may depend on a threshold effect around 10$^{12}$L$\sun$ \citep[see also][]{Tran01}.  On the other hand, it may be that the different classification schemes (V99 used the V95 scheme) are responsible for the discrepancy. We therefore re-examined the data using the V95 classifications and found that the proportion of starburst galaxies
did indeed decrease with IR luminosity, falling from $\sim$60\% at
L$_{\rm IR} < 10^{10.75}$L$\sun$ to $\sim$45\% at L$_{\rm IR}>
10^{11.25}$L$\sun$ (Table 5, Fig. 2b). No obvious trend with luminosity was seen in the Seyferts but the proportion of LINERs, borderline and
ambiguous galaxies rose from 14\% (L$_{\rm IR}<$10$^{11}$L$\sun$) to 36\% (L$_{\rm IR}>$10$^{11.25}$). A Student t-Test indicates that the mean L$_{\rm IR}$ of the combined LINER, borderline and ambiguous galaxies is higher than that of the V95 classified HII galaxies with a probability of $<$1\% of being due to chance.  Note that 21 galaxies are classified as LINERs, ambiguous or borderline galaxies under the V95 system, whereas the majority of these galaxies (15/21) are reclassified as starburst galaxies under the K01 system.  

It has long been known \citep[e.g.][]{Degrijp85} that Seyferts exhibit
warmer IR colors, as measured by the ratio of emission at 60$\mu$m
to that at 25$\mu$m.  Consequently, AGN candidates have often been
selected on the basis of warm IR colors \citep[e.g.][ K01]{Degrijp85}.
We find that Seyferts make up 20\% (9/43) of the galaxies with
S$_{60}$/S$_{25}<8$ (the selection criterion used in K01 to select AGN candidates) and only 6\% (3/50) galaxies with
S$_{60}$/S$_{25}>8$ with a probability of $<2\%$ of such a
distribution being obtained by chance, in good agreement with previous
studies.  The warm FIR colors exhibited by AGN are believed to be due
to the hot gas and dust in the torus which is heated by emission from the
accretion disk \citep{Pier93,Efstathiou95}. Using the K01 classification scheme we find that there is no significant difference between the proportion of starburst galaxies with S$_{60}$/S$_{25}<8$ (70\%, 30/43) and S$_{60}$/S$_{25}>8$ (60\%, 30/50). The majority (14/15) of galaxies for which we were unable to measure line ratios have S$_{60}$/S$_{25}>8$, confirming our suspicion that these galaxies do not possess an excess of hot dust and are not active, hence the lack of measurable H$\beta$ emission.

The galaxies classified as LINERS, borderline and ambiguous using the
V95 system tend to have cool FIR colors (14/21 have S$_{60}$/S$_{25}>8$) with a statistical probability $<$0.1\% that the difference between their mean FIR colour compared to that of the starburst galaxies is due to chance. By contrast we find that the galaxies classified as borderline using the K01 system tend to have warm FIR colors (2/2 have S$_{60}$/S$_{25}<8$) while those with an ambiguous classification tend to have cooler FIR colors (3/4 have S$_{60}$/S$_{25}>8$) and starburst-like [NII] and [SII] line strengths with unusually strong [OI] emission\footnote{Curiously the one ambiguous galaxy which exhibits warm IR colors (IRAS 00085-1223) also exhibits strong (AGN-like) [NII] emission, possibly indicating that it is a composite galaxy.}. The small number of borderline and ambiguous galaxies from the K01 system means that it is not possible to determine whether these trends are significant.  

Our results imply that for galaxies with L$_{\rm IR}<10^{12}$L$\sun$ there is no evidence that the fraction of Seyfert galaxies increases with L$_{\rm IR}$. However, there is some evidence that the proportion of galaxies classified as LINERs, borderline and ambiguous galaxies using the V95 classification system does increase with L$_{\rm IR}$. These galaxies are reclassified as starburst/ambiguous galaxies, using the K01 system, and the fact that they tend to have cool FIR colors, unlike the Seyfert galaxies, supports the view that they are ``unusual'' (metal-rich) starbursts rather than AGN. One interpretation of this is that as L$_{\rm IR}$ approaches 10$^{12}$L$\sun$ starburst galaxies
migrate into the metal-rich starburst region of the K01 diagrams,
where they would become classified as LINERs/ ambiguous in the more traditional
diagnostic diagrams (thus skewing the statistics away from starbursts).
This will be investigated further in Appleton et al.

\subsection{Radio-FIR Properties versus Galaxy Classification}

The total integrated radio luminosities of the Southern COLA galaxies at 4.8, 2.5 and 1.4GHz were published in Paper I together with the results of Australian
LBA snapshot observations. At 4.8GHz, 104 of the 107 galaxies were
detected at the 5$\sigma$ level. A summary of the radio measurements is given in Table 4.

Three galaxies (IRAS 05449-0651, IRAS 10484-0153 and IRAS 15555-6610)
were not detected at 4.8GHz.  IRAS 10484-0153 is classified as a
starburst galaxy and has a 60 micron flux of 4.51\,Jy so should have
been detected at radio wavelengths if it was to follow the radio-FIR
correlation.  The other two galaxies are predicted to be weak at radio
wavelengths based on their far-IR fluxes. IRAS 05449-0651 and IRAS
15555-6610 were originally included in the COLA sample as their
60$\mu$m fluxes were listed as greater than 4\,Jy in the {\it IRAS Point Source
Catalogue}, but their fluxes in the {\it Faint Source Catalogue} fall
well below 4\,Jy (0.85 and 2.41\,Jy respectively).  These three galaxies have
been excluded from the analysis which follows.

A tight correlation was seen between the radio and FIR luminosity of
the southern COLA galaxies ( Fig. 2; Paper I). This correlation is
well-known and is seen across many different galaxy magnitudes and types
\citep{Helou85,Sanders85,Yun01}. It is often attributed to
star formation in the disk of the galaxies: the FIR emission produced
by dust heated by young stars and the radio emission from electrons
accelerated in supernovae. Following \citet{Helou85}, we define the parameter
$q=$ Log$({\rm S_{FIR}/S_{radio}})$ to quantify the relative strengths of the radio-FIR emission, where ${\rm
S_{radio}}$ is the total radio flux of the galaxy at 4.8GHz. Table 6 shows the mean,
median and standard deviation, $\sigma$, of $q$ for the different
galaxy types as classified using the K01 classification scheme. The
median $q$ for the whole Southern COLA galaxies is 2.80$\pm$0.03. The
59 starburst galaxies (excluding IRAS 10484-0153 which was not detected at radio wavelengths) in our sample exhibit a tight radio-FIR
correlation with a median $q$ = 2.83 (mean $q$=2.81), and a relatively small scatter
($\sigma$=0.2). The Seyfert galaxies (11 galaxies as IRAS 15555-6610
is now excluded) exhibit a much larger scatter ($\sigma$=0.49) and a
significantly smaller median $q$ of 2.55 (mean $q$=2.41), indicating that the Seyfert galaxies exhibit a radio excess on the radio-FIR correlation. The
Student t-Test indicates that the difference in mean $q$ between the
two samples is significant with a probability $<$10$^{-5}$ of being a chance occurance. Since $q$ is related to the radio and FIR luminosity of the galaxies and we found no correlation between the FIR luminosity and the spectral classification of the galaxies (section 5.1), we expect to find some correlation between  radio luminosity and spectral classification. A Student t-Test on the data reveals that this is indeed the case as galaxies classified as Seyferts have a group average radio luminosity higher than that of starbursts, with a probability $P$=1.7$\times$10$^{-4}$ of such a distribution being due to chance. Note, however, that the correlation of spectral type with $q$ is much stronger than that with radio luminosity. 

The galaxies with ambiguous or borderline classifications exhibit
similar values of $q$ to the starburst galaxies (mean $q$=2.78 $\pm$0.05 and 2.83$\pm$0.05 respectively).  This result suggests
that the radio emission in these galaxies is dominated by star
formation-related processes, confirming the result of \citet{Roy98}.

\section{Compact Cores versus Spectral Classification} 

High resolution, single baseline snapshots were obtained with the Australian LBA of 105/107 of the galaxies in the Southern COLA sample. The remaining two galaxies, of which only one is included in the sub-sample discussed in this paper (IRAS 05053-0805), were not observed due to technical difficulties. These observations, described in greater detail in Paper I, were conducted at 2.3GHz and would detect radio emission $\ge 1.5$mJy from structures $\le 0.5''$ in size, corresponding to brightness temperatures $>10^{5}$K. Since we did not obtain LBA observations of IRAS 05053-0805 it has been excluded from the analysis which follows.

Compact radio cores were detected in 9/105 ($\sim$ 9\%) objects
observed and the detected fluxes (and upper limits for the non-detections) are given in Table 4. The cores detected ranged in luminosity from L=10$^{3.5-5.0}$L$\sun$ (7 have L$>$10$^{4}$L$\sun$) while the median upper limit for non-detections is L=10$^{3.78}$L$\sun$ with only 6 sources having upper limits on their core luminosity $>10^{4}$. The COLA galaxy sample was constrained to lie within a thin shell of redshift space in order to reduce  any distance dependent biases. The range exhibited in detection upper limits owed as much to differences the sensitivities of the telescopes used in the course of these observations as to the spread of distances of the objects in our sample.  

Six of the eleven galaxies (55\%) in which we detect compact radio cores are classified optically as Seyfert galaxies while only three were detected in the 59 galaxies classified as starburst by the K01 system. The positions of the galaxies with compact cores on the optical diagnostic diagrams are shown in Fig. 3 (filled symbols) from which we can see that two of the three starburst galaxies (IRAS 09375-6951 and IRAS 13097-1531) are actually classified as borderline objects using the V95 system.  

To determine whether the detection rates of compact cores in Seyferts and starburst galaxies were significantly different, we performed the following statistical analysis, assuming a binomial distribution with the detection/non-detection of a compact core as the success/failure criterion. Mindful of any biases due to e.g. different detection limits for sources, we used only those sources in which cores were detected with L$>10^{4}$L$\sun$ or in the case of non-detections, upper limits of L$<10^{4}$L$\sun$. This resulted in the elimination of 2 of the 9 sources in which compact cores were detected (one Seyfert and one starburst galaxy) and 6 of the galaxies without compact cores (4 starburst galaxies and 2 with ambiguous classifications). We further restricted our analysis to include only those galaxies for which we have obtained a spectral classification. The resultant sample contained 10 Seyferts, 53 starburst galaxies and 4 ambiguous or borderline objects. Initially the detection rate of compact cores in starburst galaxies was used to estimate the probability of detecting a compact core in a given source (2/53) assuming all galaxies have the same probability of containing a compact core, irrespective of spectral type. We found that the probability of detecting 5 cores in a sample 10 objects is $P=1.4 \times 10^{-5}$, indicating that the detection rate of cores in Seyferts is indeed significantly higher than that in starburst galaxies. Of course, the actual detection rate may be underestimated by using the starburst detection rate so our calculations were repeated for a detection rate of 7/67 (i.e. that of cores L$>10^{4}$L$\sun$ in all galaxies with a spectral classification). This time the probability of detecting 5 cores in a subset of 10 objects rose to $0.0018$ (still less than 0.5\%) while the probability of detecting only two cores out of a subset of 53 galaxies became highly unlikely (P$<10^{-5}$). It is therefore clear that Seyferts and starburst galaxies exhibit statistically different detection rates for compact radio cores.

The Seyfert galaxies with compact cores display a significant radio
excess (median $q$=2.27) relative to the starburst galaxies without compact cores (Fig. 4) with a probability $<10^{-8}$ (from the Student t-Test) that this
excess is due to chance. Again these galaxies also have higher radio luminosities than the starburst galaxies (median radio power =3.8$\times$ 10$^{22}$ W Hz$^{-1}$ and 6.6$\times$ 10$^{21}$ W Hz$^{-1}$ respectively), but the significance of this (P$\sim 10^{-7}$) is slightly smaller than for $q$. By contrast, the Seyfert galaxies without compact cores exhibit $q$ values and radio luminosities which are not significantly different from those displayed by the galaxies optically classified as starburst galaxies (median radio power = 5.5$\times$ 10$^{21}$ W Hz$^{-1}$, median $q$ = 2.79; Table 6).  These results imply that the source of the radio emission in the Seyfert galaxies without cores is the same as that of the starburst galaxies - i.e. star formation in the galactic disk \citep[e.g.][]{Kennicutt98} - in good agreement with \citet{Roy98}.  The AGN itself must contribute little to the overall radio emission in this scenario. The alternative explanation that the radio and FIR emission from the AGN without compact cores are related and scale in a similar manner as the radio and FIR emission from star formation seems somewhat improbable. 

The radio emission from the radio cores detected in the Seyfert
galaxies is too weak to account for their low $q$ values, and in Paper I
we concluded that the most probable explanation is that they exhibit a radio
excess due to extended structures (at the $\sim$ 100pc - 1kpc
scale) associated with the radio cores, e.g. small-scale jets and/or radio
lobes. Indeed one of the objects in which a compact radio core is
detected (IRAS 13197-1627) does display an extended linear structure
$\sim$280pc in length \citep{Kinney00}.  

Although we cannot rule out the presence of a compact radio core in the non-LBA detected Seyfert galaxies we can place an upper limit of L$<10^{3.7}$L$\sun$ on any such core. Only one of the Seyferts in which we detect a compact core (IRAS 12329-3938) has a core luminosity L$<10^{4}L\sun$ and while this source also exhibits the lowest FIR luminosity and second lowest radio luminosity of all the Seyferts (including those not detected with the LBA) it still exhibits a significant radio excess ($q$=2.5). The non-detection of compact cores and lack of a significant radio excess in the 5 Seyferts without cores
therefore implies that not only must their radio cores (if present) be intrinsically weaker relative to their host galaxy than those in the Seyferts with cores but also that these objects lack extended radio
structures.
 
The previous statement should not be taken to mean that the Seyfert
galaxies without compact cores are ``weak'' AGN. From the estimated
AGN contributions to the optical line ratios presented in Table 3,
there is no evidence to suggest that the Seyferts with compact cores
exhibit larger AGN contributions to the optical line ratios than those
without (Fig. 3). Similarly, Student t-Tests on the data indicate that
there are no statistically significant differences between the mean IR
luminosity and colors of the Seyferts with compact cores and those
without. In fact we argue that it is only at radio wavelengths that two Seyfert populations emerge; ``radio excess'' Seyferts, which exhibit compact radio cores and an excess of radio emission relative to their FIR luminosity, and ``radio-quiet'' Seyferts in which the vast majority of radio emission can be attributed to star formation in the host galaxy.  

Compact cores are also detected in three galaxies classified as starbursts by the K01 system. In the Seyfert galaxies, the significant radio excess we observe  suggests that their compact cores are associated with a radio jet but the origin of the compact radio emission in the starburst galaxies is not so obvious. While the presence of a high--brightness--temperature compact radio core is usually evidence of an AGN, the situation is much less certain for lower luminosity compact cores.

Significant differences do exist between previous studies of
compact cores in starburst galaxies. In the South, \citet{Norris90}
detected compact ($<0.1''$) radio cores at 2.3 GHz in only 5\% of their starburst galaxies and 32\% of AGNs with ${\rm L}_{\rm FIR} > 10^{10}{\rm
L}_{\sun}$.  With the same spatial resolution and sensitivity ($>$5\,mJy) as the Norris et al. survey, \citet{Heisler98} detected no compact cores in their
starburst galaxies and a much larger fraction (90\%) of the AGN in
their survey of 60$\mu$m peakers. \citet{Kewley00} found a larger
fraction ($\sim37\%$) of starbursts and a similar fraction (80\%) of
AGN-type galaxies with compact ($<$0.1'') cores.  Some of these differences may
be due to differing sensitivity limits, or to the different ways in which the parent samples were selected.  The majority of the Kewley et al. starbursts have core
fluxes $<5$\,mJy and would not have been detected in the Norris et al. and Heisler et al. surveys.  As mentioned in the introduction, SLL98 also carried out a VLBI study in the north and found little difference in
detection rate ($53-54$\%) between optically classified AGN and
starbursts \citep[see also][]{Lonsdale93}.  The SLL98 observations
generally have a higher resolution than the southern observations, and
their sample contains galaxies with higher luminosities (${\rm L}_{\rm
FIR} > 10^{11.25}{\rm L}_{\sun}$) -- both factors that could account
for the different statistics. In all the surveys, the authors were unable to rule out a SNRe origin for the low luminosity compact radio emission detected in starburst galaxies.

Our results confirm that compact cores tend to be more prevalent in
galaxies with Seyfert classifications (6/11) than starburst galaxies
(3/58). Of the three starburst galaxies, two of them, IRAS 13097-1531
and IRAS 09375-6951, have core radio luminosities L$>10^{4}$\,L$\sun$,
which has been taken by \citet{Kewley00} to indicate a likely AGN
rather than a SNR origin. Despite the fact that these galaxies have
cool IRAS colors (see earlier discussion), they also
exhibit a slight excess in the radio-FIR correlation.  

Notes on the individual Seyfert galaxies in our sample and the starburst galaxies in which compact cores were detected are given in the Appendix.

\section{Conclusions}

We present optical spectroscopic observations of 93 galaxies selected
from the southern portion of the COLA survey.  The spectra have been
used to classify the galaxies into various types based on the strength
of key emission lines using both traditional semi-empirical (V95) and a
new theoretically predicted (K01) diagnostic ratios. These spectroscopic diagnostics of nuclear activity are compared with
other probes, such as IRAS colors and the existence of a compact radio
core as a measure of activity. The following conclusions are reached:

\begin{itemize}

\item For the COLA (south) sample, we find that the fraction of Seyfert
galaxies in the sample is independent of the diagnostic method used
(16\%), but that the fraction of starburst systems does depend on the
choice of diagnostic. Using the V95 method, we find $\sim$57\% are starbursts, 8\% are LINERs and 19\% are ambiguous or borderline, whereas under the K01 system, 77\% are
classified as starbursts, none are classified as LINERs and a smaller fraction (8\%) fall into the
ambiguous or borderline category. This is a result of model
predictions which suggest that metal-rich starbursts can have
emission-line ratios which place them into the phase-space occupied  by LINERs 
in the semi-empirical diagnostic diagrams.

\item The COLA sample covers only the far-IR luminosity range of normal galaxies to Luminous Infrared Galaxies (LIRGs) and falls short of sampling the ULIRG
region (L$_{\rm IR} > 10^{12} L\sun$). Using the K01 classification system, our spectra suggest little change in the fraction of Seyfert galaxies relative to starbursts as a function of far-IR luminosity --in contrast to previous work. We note, however, that the narrow range of luminosities covered by our sample may not allow us to identify a weak trend in our data. Our result suggests either that there is a threshold IR luminosity above which AGNs begin to dominate the population, or that, as the
galaxies become more IR luminous, there is a tendency for them to
migrate into the metal-rich starburst region of the K01 diagrams,
where they would become classified as LINERs in the more traditional
diagnostic diagrams (thus skewing the statistics away from starbursts).  We
note that the COLA sample contains a high fraction of interacting
systems which are known to be metal-rich in their nuclei; such a
migration would be consistent with galaxies becoming IR luminous as
they evolved through collisions.

\item In good agreement with previous studies we find that the proportion
of Seyfert galaxies in our sample increases with warmer FIR colors
(S$_{60}$/S$_{25}$) and that Seyfert galaxies emit more flux at 25$\mu$m
 than starburst galaxies with the same IR luminosity.

\item Comparison of the optical spectral types of the galaxies with the
radio-FIR observations reveals that the galaxies classified optically
as starbursts exhibit a tight radio-FIR correlation with little
scatter and a median $q$ =2.83. The galaxies classified
optically as Seyferts exhibit a slight (but statistically significant) radio excess on the radio-FIR correlation with a median $q$ =2.55 and a somewhat larger scatter.

\item Compact radio cores were detected in 6/11 ($\sim$55\%) of galaxies classified optically as Seyferts and 3/59 ($\sim$ 5\%) of starburst galaxies. 
The upper limits on the detection of the compact cores in the remaining Seyferts (5/11) are between 10$^{3.5}$ to $10^{3.7}$ L$\sun$ indicating that their compact cores, if present, must be at the lower end of the luminosities detected (10$^{3.4}$--10$^{5.6}$ L$\sun$). 

\item All the Seyfert galaxies in which compact cores were detected lie significantly above the mean radio-FIR correlation  with a median $q$=2.26. This radio excess is attributed to pc- to kpc-scale radio
structures, because the compact radio emission is only a small part of
the observed excess (see Paper I). The Seyfert galaxies without compact cores do not exhibit the same radio excess as the Seyfert galaxies with compact cores and have a similar median $q$ (= 2.79) to that of the starburst galaxies. This suggests the presence of two different kinds of Seyferts, ``radio-excess'' Seyferts (these galaxies are not formally radio loud), which exhibit radio cores and a large-scale radio structure, and ``radio-quiet''
Seyferts, with low luminosity radio cores and a radio continuum dominated by star formation in the host galaxy. We emphasize that both the {\em optical strength} of
the AGN signature seen in the line diagnostic diagrams and the FIR color of the Seyfert galaxies is insensitive
to the difference -- the radio quiet Seyfert galaxies often have very
dominant AGN-signatures in their optical spectra. We are currently
investigating what factors might lead to these differences in the
properties of the host galaxies.

\item It is not clear whether the compact cores detected in the three galaxies classified optically as starbursts are obscured AGN or complexes of RSNe. They do not exhibit a significant radio excess and the luminosities of the compact cores detected (10$^{3.75}$ to $10^{4.25}$ L$\sun$) are consistent with either origin. This degeneracy may be resolved by VLBI or X-ray observations which could reveal the presence of a collimated pc-scale jet or a hard X-ray continuum if the cores are indeed AGN.  

\item Galaxies classified as ambiguous and borderline in the K01 scheme have little
discernible radio excess (median $q$=2.78, 2.83 respectively), suggesting that they have
more in common with starbursts than AGN. Since these galaxies would
have been classified as LINERs on the V95 scheme, this would seem to
add support to the idea that some LINER objects are metal-rich
starbursts.

\end{itemize}

\section{Acknowledgments}
This paper is dedicated to Charlene Anne Heisler (deceased), who was 
the driving force behind this project for many years. We thank I. Wormleaton for her help with the reduction of the optical spectroscopic data. We thank S. Veilleux for his careful reading of this paper and his suggestions which have greatly improved its clarity.
This research has made use of the NASA/IPAC Extragalactic Database
(NED) which is operated by the Jet Propulsion Laboratory, California
Institute of Technology, under contract with the National Aeronautics
and Space Administration. It has also has made use of NASA's
Astrophysics Data System Bibliographic Services and of the on-line
NRAO VLA Sky Survey database (Condon et al. 1998). EAC is supported by the Anglo-Australian Fellowship and LK is
supported by a Harvard-Smithsonian CfA Fellowship.  MAD
acknowledges the support of the Australian National University and the
Australian Research Council through his ARC Australian Federation
Fellowship, and under the ARC Discovery project
DP0208445. VC would like to acknowledge the partial
support of JPL contract 960803. AZ acknowledges partial support from the Chandra grants G01-2116X and G02-3150X.
}

\begin{deluxetable}{lcccccccccccccccc}
\tabletypesize{\scriptsize}
\rotate
\setlength{\tabcolsep}{0.03in} 
\tablewidth{0pc} 
\tablecolumns{17}
\tablecaption{The observed line ratios and optical classifications for the COLA sample.}
\tablecomments{ Column 2 identifies the aperture for those galaxies which exhibited multiple ``hotspots''. Column 3 indicates whether the galaxy spectrum was obtained as part of this work (1) or K01 (2). The logarithm of the line ratios is given in columns 4 and 6-9. The values quoted in column 4 are before the data were corrected for extinction along the line of sight whereas the line ratios quoted in columns 6-9 are extinction corrected. The errors in the line ratios vary from object to object but typical errors are $\pm$ 0.017 dex for Log([OIII]/H$\beta$), $\pm$ 0.005 dex for Log([NII]/H$\alpha$), 0.005dex for Log([SII]/H$\alpha$) and $\pm$ 0.04dex for Log([OI]/H$\alpha$) to be $\pm$ 0.04dex. The optical classifications of the galaxies from the V95 scheme are given in columns 10-12 and those using the K01 scheme are shown in columns 14-17. Classifications are as follows: Sy=Seyfert, HII=starburst, AMB=ambiguous classification, B=galaxy lies within 0.5 dex of the AGN/HII partition, AGN=AGN (probably Seyfert), L=LINER. See text for more details. Notes are as follows: $^{1}$ All ``narrow'' emission lines unusually broad (FWHM $\sim$ 1000 kms$^-1$); $^{2}$ Broad H$\alpha$ and H$\beta$ detected; $^{3}$ Additional broad H$\alpha$ component detected (FWHM $\sim$ 2000 kms$^{-1}$), but no additional broad H$\beta$ component.}
\tablehead{
\colhead{IRAS Name} & \colhead{ap.}& \colhead{Ref}& \colhead{$\frac {H\alpha}{H\beta}$}& \colhead{E(B-V)}& \colhead{$\frac{[OIII]}{H\beta}$}&\colhead{$\frac {[NII]}{H\alpha}$}&\colhead{$\frac {[SII]}{H\alpha}$}& \colhead{$\frac {[OI]}{H\alpha}$}&\colhead{V95 $\frac {[NII]}{H\alpha}$}& \colhead{V95 $\frac {[SII]}{H\alpha}$} & \colhead{V95 $\frac {[OI]}{H\alpha}$} & \colhead{V95 Final Class} & \colhead{K01 $\frac {[NII]}{H\alpha}$}& \colhead{K01 $\frac {[SII]}{H\alpha}$}  &\colhead{K01 $\frac {[OI]}{H\alpha}$}& \colhead{K01 Final Class}\\
\colhead{(1)}& \colhead{(2)}& \colhead{(3)}& \colhead{(4)}& \colhead{(5)}& \colhead{(6)}& \colhead{(7)}& \colhead{(8)}& \colhead{(9)}& \colhead{(10)}& \colhead{(11)}& \colhead{(12)}& \colhead{(13)}& \colhead{(14)}& \colhead{(15)}& \colhead{(16)}& \colhead{(17)}\\
}
\startdata
00085-1223 &         & 1 &   1.12 &      1.45   &    0.01  &   0.10  &   -0.49  &  -1.32  &    L    &   HII   &   HII  &     AMB   &    AGN  &    HII   &   HII   &   AMB\\
00344-3349 &      1  & 2 &        & ...         &   0.50   &  -0.79  &  -0.87   & -1.74   &  HII    &  HII    &  HII   &    HII    &   HII   &   HII    &  HII    &  HII\\
00344-3349 &      2  & 2 &        & ...         &   0.48   &  -0.75  &  -0.77   & -1.67   &  ...    &  ...    &  ...   &    HII    &   HII   &   HII    &  HII    &  HII\\
00402-2350 &         & 1 &   0.61 &      0.26   &   -0.23  &   -0.10 & ...      &  ...    &    L    &  ...    &  ...   &      L    &    HII  & ...      &  ...    &   HII\\
01053-1746 &         & 1 &   0.59 &      0.31   &    0.08  &   -0.53 & ...      &  ...    &   HII   &  ...    &  ...   &     HII   &    HII  & ...      &  ...    &   HII\\
01159-4443 &      2  & 1 &   0.89 &      0.91   &    0.26  &   -0.34 & ...      &  -0.94  &  B(HII) &   ...   &    L   &      B    &    HII  & ...      &   AGN   &   AMB\\
01159-4443 &      1  & 1 &   1.13 &      1.56   &   -0.17  &   -0.30 & ...      &  -1.43  &   HII   &   ..    &   HII  &     HII   &    HII  & ...      &   HII   &   HII\\
01165-1719 &         & 1 &   0.87 &      0.96   &   -0.20  &   -0.47 &   -0.58  &  -1.53  &   HII   &   HII   &   HII  &     HII   &    HII  &    HII   &   HII   &   HII\\
01326-3623 &         & 1 &   ...  &       ...   & ...      &    ...  & ...      &  ...    &  ...    &   ...   &   ...  &   No lines&    ...  & ...      &     ... &    No lines\\
01341-3734 &      1  & 2 &  0.67  &     0.41    &  -0.39   &  -0.29  &  -0.65   & -1.63   &  HII    &  HII    &  HII   &    HII    &   HII   &   HII    &  HII    &  HII\\
01341-3734 &      2  & 2 &  0.57  &     0.17    &  -0.20   &  -0.30  &  -0.66   & -1.63   &  HII    &  HII    &  HII   &    HII    &   HII   &   HII    &  HII    &  HII\\
01384-7515 &         & 1 &   0.88 &      0.97   &   -0.33  &   -0.28 &   -0.55  &  -1.42  &   HII   &   HII   &   HII  &     HII   &    HII  &    HII   &   HII   &   HII\\
02015-2333 &         & 2 &  0.79  &     0.69    &  -0.63   &  -0.28  &  -0.60   & -1.62   &  HII    &  HII    &  HII   &    HII    &   HII   &   HII    &  HII    &  HII\\
02069-1022 &         & 1 &   0.68 &      0.43   &   -0.17  &   -0.03 &   -0.42  &         &    L    &  B(HII) &   ...  &      B    &  B(HII) &   HII    &  ...    &   HII\\
02072-1025 &         & 1 &   1.10 &      1.40   &    0.13  &   -0.23 &   -0.49  &  -1.24  &   B(A)  &  B(HII) &   B(A) &      B    &    HII  &    HII   &  B(HII) &   HII\\
02140-1134 &         & 1 &   0.83 &      0.87   &   -0.44  &   -0.31 &   -0.65  &         &   HII   &   HII   &  ...   &     HII   &    HII  &    HII   &  ...    &   HII\\
02281-0309 &         & 1 &        & ...         & ...      &   0.14  & ...      &         &         &         &   ...  &    H abs  &   ...   & ...      &  ...    &   H abs\\
02433-1534 &         & 1 &   1.04 &      1.28   &   -0.18  &   -0.33 &   -0.54  &  -1.05  &   HII   &   HII   &    L   &     AMB   &    HII  &    HII   &   B(A)  &   HII\\
02436-5556 &         & 1 &        & ...         & ...      &   -0.13 &   -0.96  &   ...   & ...     &  ...    & ...    &      N    &  ...    & ...      &  ...    &    No lines\\
02476-3858 &         & 1 &        & ...         & ...      &   ...   & ...      &    ...  &   ...   &   ...   &   ...  &      N    &         & ...      &   ...   &    No lines\\
03022-1232 &         & 1 &   1.26 &      1.77   &    0.48  &   -0.13 &   -0.45  &         &   AGN   &   AGN   &        &      Sy   &    AGN  &    AGN   &  ...    &   AGN\\
04118-3207 &         & 1 &   0.66 &      0.38   &    0.44  &   -0.12 &   -0.54  &  -1.40  &    L    &  B(HII) &  B(HII)&     B     &    AGN  &    B(A)  &   B(A)  &    B\\
04210-4042 &         & 1 &   0.72 &      0.60   &   -0.66  &   -0.30 &   -0.70  &  -1.59  &   HII   &   HII   &   HII  &     HII   &    HII  &    HII   &   HII   &   HII\\
04315-0840 &         & 2 &  1.11  &     1.43    &  -0.09   &  -0.22  &  -0.64   & -1.76   &  B(A)   &  HII    &  HII   &    HII    &   HII   & ...      &  HII    &  HII\\
04335-2514 &         & 1 &        & ...         & ...      &   0.04  &   -0.37  &         &  ...    &  ...    &  ...   &    H abs  &  ...    & ...      &  ...    &  H abs\\
04370-2416 &         & 1 &   0.59 &      0.32   &   -0.61  &   -0.47 &   -0.57  &  -1.57  &   HII   &   HII   &   HII  &     HII   &    HII  &    HII   &   HII   &   HII\\
04461-0624 &         & 1 &   0.45 &      0.00   &    1.08  &   0.52  &    0.27  &  -0.28  &   AGN   &   AGN   &   AGN  &    Sy2$^{1}$    &    AGN  & ...      &   AGN   &   AGN$^{1}$\\
04501-3304 &         & 1 &   0.99 &      1.15   &   -0.57  &   -0.14 &   -0.79  &         &    L    &   HII   &        &     AMB   &    HII  &    HII   & ...     &   HII\\
04558-0751 &         & 1 &        & ...         & ...      &   -0.05 &   -0.60  &  ...    &  ...    &  ...    &  ...   &  No Lines & ...     & ...      & ...     &  No lines\\
04569-0756 &         & 1 &        & ...         & ...      &   -0.15 &   -0.46  &         &         &         &        &    H abs  &  ...    & ...      &  ...    &  H abs\\
04591-0419 &         & 1 &   0.42 &      0.00   &    0.45  &   -1.03 &   -0.87  &  -1.96  &   HII   &   HII   &   HII  &     HII   &    HII  &    HII   &   HII   &   HII\\
04595-1813 &         & 1 &   0.69 &      0.54   &   -0.10  &   -0.49 &   -0.51  &         &   HII   &   HII   &        &     HII   &    HII  &    HII   &  ...    &   HII\\
05041-4938 &         & 1 &   0.84 &      0.89   &   -0.38  &   -0.51 &   -0.73  &         &   HII   &   HII   &        &     HII   &    HII  &    HII   &  ...    &   HII\\
05053-0805 &         & 2 &  0.95  &     1.06    &  -0.73   &  -0.31  &  -0.61   & -1.64   &  HII    &  HII    &  HII   &    HII    &   HII   &   HII    &  HII    &  HII\\
05140-6213 &         & 1 &   0.80 &      0.79   &   -0.14  &   -0.37 &   -0.67  &         &   HII   &   HII   &   HII  &     HII   &    HII  &    HII   &  ...    &   HII\\
05449-0651 &         & 1 &        & ...         & ...      &   -0.39 & ...      & ...     &  ...    &   ...   &        &  No lines &         & ...      &  ...    &    No lines\\
05562-6933 &         & 1 &   0.69 &      0.53   &   -0.54  &   -0.29 &   -0.61  &  -1.58  &   HII   &   HII   &   HII  &     HII   &    HII  &    HII   &   HII   &   HII\\
06295-1735 &         & 1 &   0.63 &      0.40   &   -0.35  &   -0.38 &   -0.58  &  -1.49  &   HII   &   HII   &   HII  &     HII   &    HII  &    HII   &   HII   &   HII\\
06592-6313 &         & 2 &  1.07  &     1.34    &  -0.60   &  -0.10  &  -0.61   & -1.39   &   L     &  HII    &  HII   &    AMB    &   HII   &   HII    &  HII    &  HII\\
08175-1433 &         & 1 &   0.67 &      0.49   & ...      &   -0.40 &   -0.78  &  -1.96  & ...     &  ...    &   ...  & No lines  &   ...   & ...      & ...     &    No lines\\
08225-6936 &         & 1 &   0.80 &      0.80   &   -0.45  &   -0.30 &   -0.62  &         &   HII   &   HII   &   ...  &      H    &    HII  &    HII   &  ...    &   HII\\
08364-1430 &         & 1 &   0.70 &      0.56   &   -0.43  &   -0.33 &   -0.44  &  -1.53  &   HII   &  B(HII) &   HII  &     HII   &    HII  &    HII   &   HII   &   HII\\
08438-1510 &         & 2 &  1.11  &     1.42    &   0.15   &  -0.17  &  -0.38   & -1.11   &   L     &   L     &   L    &     L     &  B(HII) &  B(HII)  &  AGN    &   B\\
09006-6404 &         & 1 &   0.72 &      0.61   &    0.28  &   -0.41 &   -0.53  &  -1.43  &   HII   &  B(HII) &   HII  &     HII   &    HII  &    HII   &   HII   &   HII\\
09248-1918 &         & 2 &  0.92  &     0.98    &  -0.42   &  -0.21  &  -0.51   & -1.47   &  B(A)   &  HII    &  HII   &    HII    &   HII   &   HII    &  HII    &  HII\\
09375-6951 &      1  & 1 &   0.98 &      1.14   &   -0.22  &   -0.26 &   -0.42  &         &  B(HII) &  B(HII) &        &      B    &    HII  &    HII   &         &   HII\\
09375-6951 &      2  & 1 &   0.75 &      0.67   &   -0.50  &   -0.36 &   -0.51  &  -1.19  &   HII   &   HII   &  B(HII) &    HII   &    HII  &    HII   &   HII   &   HII\\
10015-0614 &      1  & 1 &   0.81 &      0.82   &   -0.59  &   -0.37 &   -0.70  &  -1.59  &   HII   &   HII   &   HII  &     HII   &    HII  &    HII   &   HII   &   HII\\
10015-0614 &      2  & 1 &   0.59 &      0.30   &   -0.36  &   -0.49 &   -0.64  &  -1.78  &   HII   &   HII   &   HII  &     HII   &    HII  &    HII   &   HII   &   HII\\
10221-2317 &         & 1 &   0.80 &      0.72   &    0.13  &   -0.09 &   -0.46  &  -1.33  &    L    &  B(HII) &   HII  &      B    &    B(A) &    HII   &   HII   &   HII\\
10409-4557 &         & 1 &   0.68 &      0.44   &   -0.03  &   -0.11 & ...      &  ...    &    L    &   ...   &   ...  &      L    &    HII  & ...      &   ...   &   HII\\
10484-0153 &         & 1 &   0.52 &      0.16   &   -0.88  &   -0.36 &   -0.48  &  -1.46  &   HII   &   HII   &   HII  &     HII   &    HII  &    HII   &   HII   &   HII\\
10567-4310 &         & 1 &   0.95 &      1.14   &$<$-0.77  &   -0.31 &   -0.73  &  -1.67  &   HII   &   HII   &   HII  &   No OIII &    HII  &    HII   &   HII   & No OIII\\
11005-1601 &         & 1 &   0.77 &      0.72   &   -0.72  &   -0.35 &   -0.56  &  -1.71  &   HII   &   HII   &   HII  &     HII   &    HII  &    HII   &   HII   &   HII\\
11254-4120 &         & 1 &   0.87 &      0.87   &   -0.44  &   -0.04 &   -0.57  &  -1.33  &    L    &   HII   &   HII  &     AMB   &    HII  &    HII   &   HII   &   HII\\
11328-4844 &         & 1 &        & ...         & ...      &   -0.11 &   -0.38  &  -1.06  & ...     &    ...  &   ...  &    H abs  &         & ...      &         &  H abs\\
11409-1631 &         & 2 &  0.58  &     0.20    &  -0.15   &  -0.59  &  -0.54   & -1.70   &  HII    &  HII    &  HII   &    HII    &   HII   &   HII    &  HII    &  HII\\
12042-3140 &      1  & 1 &   0.43 &      0.00   &   -0.05  &   -0.25 & ...      &  -0.33  &  B(HII) &  ...    &    L   &      B    &    HII  & ...      &    L    &   AMB\\
12042-3140 &      2  & 1 &   0.90 &      1.02   &   -0.27  &   -0.44 & ...      &   ...   &   HII   &  ...    &  ...   &     HII   &    HII  & ...      &  ...    &   HII\\
12112-4659 &         & 1 &   0.92 &      1.08   &   -0.52  &   -0.30 &   -0.66  &  -1.90  &   HII   &   HII   &   HII  &     HII   &    HII  &    HII   &   HII   &   HII\\
12115-4657 &         & 1 &   0.72 &      0.60   &    0.04  &   -0.31 &   -0.60  &  -1.78  &   HII   &   HII   &   HII  &     HII   &    HII  &    HII   &   HII   &   HII\\
12120-1118 &         & 1 &   0.71 &      0.60   &   -0.50  &   -0.39 &   -0.53  &  -1.52  &   HII   &   HII   &   HII  &     HII   &    HII  &    HII   &   HII   &   HII\\
12171-1156 &         & 1 &   0.73 &      0.64   &   -0.72  &   -0.35 &   -0.60  &  ...    &   HII   &   HII   &  ...   &     HII   &    HII  &    HII   &   ...   &   HII\\
12286-2600 &         & 2 &  0.89  &     0.93    &  -0.64   &  -0.24  &  -0.66   & -1.64   & B(HII)  &  HII    &  HII   &     HII   &   HII   &   HII    &  HII    &  HII\\
12329-3938 &         & 1 &   0.47 &    0        &    0.93  &   -0.29 &   -0.43  &  -0.93  &   AGN   &   AGN   &   AGN  &     Sy2   &    AGN  &    AGN   &   AGN   &   AGN\\
12351-4015 &      1  & 1 &   0.65 &      0.37   &   -0.22  &   -0.51 &   -0.57  &  -0.26  &   HII   &   HII   &    L   &     AMB   &    HII  &    HII   &    L    &   AMB\\
12351-4015 &      3  & 1 &   0.70 &      0.47   &   -0.60  &   -0.28 &   -0.50  &  -0.95  &   HII   &   HII   &    L   &     AMB   &    HII  &    HII   &   B(A)  &   HII\\
12351-4015 &      2  & 1 &   0.73 &      0.63   &   -0.61  &   -0.50 &   -0.57  &  -1.36  &   HII   &   HII   &   HII  &     HII   &    HII  &    HII   &   HII   &   HII\\
12596-1529 &      1  & 2 &  0.97  &     1.10    &  -0.35   &  -0.28  &  -0.62   & -1.49   &  HII    &  HII    &  HII   &    HII    &   HII   &   HII    &  HII    &  HII\\
12596-1529 &      2  & 2 &  0.79  &     0.70    &  -0.15   &  -0.49  &  -0.65   & -1.85   &  HII    &  HII    &  HII   &    HII    &   HII   &   HII    &  HII    &  HII\\
12596-1529 &      3  & 2 &  0.72  &     0.53    &   0.10   &  -0.68  &  -0.55   & -1.48   &  HII    &  HII    &  HII   &    HII    &   HII   &   HII    &  HII    &  HII\\
13001-2339 &         & 1 &   0.83 &      0.78   &   -0.11  &   -0.13 & ...      &  -0.84  &    L    &  ...    &    L   &      L    &    HII  & ...      &   AGN   &   AMB\\
13035-4008 &         & 1 &   0.59 &      0.22   &    0.75  &   0.15  &   -0.01  &  ...    &   AGN   &  ...    &  ...   & Sy1$^{2}$ &    AGN  &    AGN   &   ...   &   AGN\\
13097-1531 &         & 1 &   0.61 &      0.28   &   -0.46  &   -0.20 &   -0.45  &  -1.25  &   B(A)  &  B(HII) &   HII  &      B    &    HII  &    HII   &   HII   &   HII\\
13135-2801 &         & 1 &   0.79 &      0.77   &   -0.58  &   -0.30 &   -0.67  &  ...    &   HII   &   HII   &  ...   &     HII   &    HII  &    HII   &  ...    &   HII\\
13192-5208 &         & 1 &   0.82 &      0.84   &   -0.63  &   -0.32 &   -0.66  &  ...    &   HII   &   HII   &  ...   &     HII   &    HII  &    HII   &  ...    &   HII\\
13197-1627 &         & 1 &   ...  & ...         &    1.30  &   0.69  &    0.51  &   0.23  &   AGN   &   AGN   &   AGN  & Sy1.8$^{3}$&    AGN &    AGN   &   AGN   &   AGN\\
13229-2934 &         & 2 &  0.71  &     0.51    &   0.69   &  -0.04  &  -0.45   & -1.19   &   AGN   &  AGN    &  AGN   &    Sy2    &   AGN   &   AGN    &  AGN    &  AGN\\
14544-4255 &         & 1 &   0.64 &      0.35   &    0.81  &   -0.14 &   -0.23  &  -0.93  &   AGN   &   AGN   &   AGN  &     Sy2   &    AGN  &    AGN   &   AGN   &   AGN\\
14566-1629 &         & 1 &   1.01 &      1.20   &    0.57  &   0.08  &   -0.23  &         &   AGN   &   AGN   &  ...   &     Sy2   &    AGN  &    AGN   &   ...   &   AGN\\
15555-6610 &         & 1 &   0.76 &      0.62   &    1.10  &   -0.01 &   -0.34  &  -0.85  &   AGN   &   AGN   &   AGN  &     Sy2   &    AGN  &    AGN   &   AGN   &   AGN\\
16153-7001 &         & 1 &        & ...         & ...      &         & ...      &   ...   &   ...   &   ...   &   ...  & No lines  &   ...   & ...      &   ...   &    No lines\\
16229-6640 &      1  & 1 &   1.25 &      1.76   &   -0.09  &   -0.15 & ...      &  -1.32  &    L    &   ...   &   HII  &     AMB   &    HII  &    HII   &   HII   &   HII\\
16229-6640 &      2  & 1 &   0.88 &      0.97   &   -0.52  &   -0.28 & ...      &         &   HII   &   HII   &   ...  &     HII   &    HII  & ...      &   ...   &   HII\\
17138-1017 &         & 2 &  1.41  &     2.12    &  -0.34   &  -0.29  &  -0.54   & -1.37   &  HII    &  HII    &  HII   &    HII    &   HII   &   HII    &  HII    &  HII\\
17182-7353 &         & 1 &  ...   & ...         & ...      &  ...    & ...      &  ...    &  ...    &   ...   & ...    &      N    &  ...    &	  ...   & ...     &   N \\
17260-7622 &         & 1 &   0.82 &      0.83   &   -0.18  &   -0.45 &   -0.50  &         &   HII   &   HII   &   ...  &     HII   &    HII  &    HII   &  ...    &   HII\\
18093-5744 &         & 1 &   0.68 &      0.52   &   -0.28  &   -0.35 &   -0.53  &  -1.52  &   HII   &   HII   &   HII  &     HII   &    HII  &    HII   &   HII   &   HII\\
18293-3413 &         & 1 &   1.25 &      1.82   & ...      &   -0.31 & ...      &  -1.32  &  ...    &   ...   &    ... &      N    &  ...    & ...      &  ...    &    No lines\\
18341-5732 &         & 1 &   1.19 &      1.61   &   -0.24  &   -0.12 &   -0.57  &  -1.44  &    L    &   HII   &   HII  &     AMB   &    HII  &    HII   &   HII   &   HII\\
18421-5049 &         & 1 &   0.13 &    0        &    0.08  &   -0.12 &   -0.46  &         &    L    &  B(HII) &        &      B    &  B(HII) &   HII    &  ...    &   HII\\
18429-6312 &         & 2 &  0.86  &     0.85    &   1.12   &  -0.09  &  -0.38   & -1.00   &  AGN    &  AGN    &  AGN   &    Sy2    &   AGN   &   AGN    &  AGN    &  AGN\\
19543-3804 &         & 1 &   1.09 &      1.38   &    0.99  &   0.14  &   -0.26  &  -0.74  &   AGN   &   AGN   &   AGN  &      Sy   &    AGN  &    AGN   &   AGN   &   AGN\\
20305-0211 &      2  & 1 &   0.69 &      0.53   &   -0.28  &   -0.53 &   -0.64  &  -1.67  &   HII   &   HII   &   HII  &     HII   &    HII  &    HII   &   HII   &   HII\\
20305-0211 &      1  & 1 &   0.77 &      0.72   &   -0.08  &   -0.50 &   -0.76  &  -1.30  &   HII   &   HII   &   HII  &     HII   &    HII  &    HII   &   HII   &   HII\\
20309-1132 &         & 1 &   0.75 &      0.67   &   -0.48  &   -0.45 &   -0.51  &   ...   &   HII   &   HII   &  ...   &     HII   &    HII  &    HII   &   ...   &   HII\\
20486-4857 &         & 1 &   ...  & ...         & ...      &   0.14  & ...      &   ...   &   ...   &   ...   &  ...   &    H abs  &   ...   & ...      &   ...   &  H abs\\
21008-4347 &         & 1 &   0.13 & ...         &   -0.38  &   -0.41 & ...      &  ...    &   HII   &  ...    &  ...   &     HII   &    HII  & ...      &   ...   &   HII\\
21314-4102 &         & 1 &   0.81 &      0.73   &   -0.29  &   -0.12 & ...      &  -0.84  &    L    &   ...   &    L   &      L    &    HII  & ...      &   AGN   &   AMB\\
21330-3846 &      2  & 2 &  1.10  &     1.41    &  -0.24   &  -0.37  &  -0.65   & -1.21   &  HII    &  HII    & B(HII) &    HII    &   HII   &   HII    &  HII    &  HII\\
21330-3846 &      1  & 2 &  0.83  &     0.78    &  -0.18   &  -0.17  &  -0.30   &  ...    &   L     &   L     &  ...   &     L     &   HII   &   HII    &   ...   &  HII\\
21453-3511 &         & 1 &   0.79 &      0.69   &    0.80  &   -0.17 &   -0.45  &  -0.86  &   AGN   &   AGN   &   AGN  &      Sy   &    AGN  &    AGN   &   AGN   &   AGN\\
22115-3013 &         & 1 &   0.77 &      0.73   &   -0.48  &   -0.36 &   -0.55  &   ...   &   HII   &   HII   &   ...  &     HII   &    HII  &    HII   &   ...   &   HII\\
22118-2742 &         & 1 &   0.79 &      0.77   &   -0.19  &   -0.44 & ...      &   ...   &   HII   &   ...   &   ...  &     HII   &    HII  & ...      &   ...   &   HII\\
22179-2455 &         & 1 &   0.81 &  0.82       & ...      &   -0.26 &   -0.51  &   ...   &   ...   &  ...    &   ...  & No lines  &    ...  & ...      &   ...   &    No lines\\
23394-0353 &         & 1 &   0.92 &      1.08   &   -0.35  &   -0.36 &   -0.61  &  -1.36  &   HII   &   HII   &   HII  &     HII   &    HII  &    HII   &   HII   &   HII\\
\enddata 
\end{deluxetable}

\begin{deluxetable}{lcccc}
\tabletypesize{\footnotesize}
\tablewidth{0pc}
\tablecolumns{5}
\tablecaption{The classification statistics for the COLA galaxies using the V95 and K01 schemes. }
\tablecomments{``Nuclear'' classification refers to the classification of the nuclear spectrum of each galaxy. ``All apertures'' includes additional spectra from ``hotspots'' detected in some of the galaxies. }
\tablehead{
\colhead{Class}& \colhead{V95 Nuclear}&\colhead{V95 All Ap.}&\colhead{K01 Nuclear}&\colhead{K01 All Ap.}\\ 
}
\startdata
HII           & 44& 54& 59 & 70\\
AGN           & 12& 12& 12 & 12\\
LINERs        &  6&  6&  0 &  0\\
B             &  8&  9&  2 &  2\\
AMB           &  7&  9&  4 &  6\\
Not classified& 15& 15& 15 & 15\\
Total         & 93&106& 93 &106\\
\enddata 
\end{deluxetable}

\begin{deluxetable}{lccccccc}
\tabletypesize{\footnotesize}
\tablewidth{0pc}
\tablecolumns{8}
\tablecaption{The minimum and maximum percentage contribution to the observed
emission line ratios due to ionization of the gas by an AGN continuum.
This is calculated using the theoretical ionization models of K01 with
ionization by starburst activity accounting for the remainder of the line
ratios (see discussion in text for further details).}
\tablecomments{Galaxies which are classed as Sy1 or Sy2 in both the V95 and K01
classification schemes are marked with an asterix. }
\tablehead{
\colhead{Names}& \colhead{ap.}&\colhead{[NII]/H$\alpha$}&\colhead{[SII]/H$\alpha$}&\colhead{[OI]/H$\alpha$} & \colhead{Mean}&\colhead{Min} & \colhead{Max}\\
\colhead{}& \colhead{}& \colhead{(\%)}& \colhead{(\%)}& \colhead{(\%)}& \colhead{(\%)}& \colhead{(\%)}& \colhead{(\%)}\\
}
\startdata
00085-1223 &       &  15-20  &    0-15  &    0-15     & 11    &     0  &      20\\
00344-3349 &    1  &  10-10  &    0-10  &    0-5      & 4     &    0   &     10\\
00344-3349 &    2  &         &          &             &       &        &      \\
00402-2350 &       &   0-10  &    0-10  &             & 5     &    0   &     10\\
01053-1746 &       &   0-5   &    0-15  &             & 5     &    0   &     15\\
01159-4443 &    2  &   0-10  &    0-10  &    0-10     & 5     &    0   &     10\\
01159-4443 &    1  &   0-15  &    0-25  &    20-30    &15     &    0   &     30\\
01165-1719 &       &   0-5   &    0-10  &    0-10     & 6     &    0   &     10\\
01326-3623 &       &         &          &             &       &        &       \\
01341-3734 &    1  &   0-10  &    0-10  &    0-5      & 4     &    0   &     10\\
01341-3734 &    2  &   0-10  &    0-10  &    0-10     & 5     &    0   &     10\\
01384-7515 &       &   0-5   &    0-10  &    0-10     & 6     &    0   &     10\\
02015-2333 &       &   0-5   &    0-5   &    0-5      & 3     &    0   &      5\\
02069-1022 &       &   0-10  &    0-10  &             & 5     &    0   &     10\\
02072-1025 &       &   0-15  &    0-20  &    0-20     & 9     &    0   &     20\\
02140-1134 &       &   0-5   &    0-10  &             & 4     &    0   &     10\\
02281-0309 &       &         &          &             &       &        &     \\
02433-1534 &       &   0-10  &    0-5   &    0-10     & 6     &    0   &     10\\
02436-5556 &       &         &          &             &       &        &      \\
02476-3858 &       &         &          &             &       &        &     \\
03022-1232$^{*}$ & &  10-30  &   10-50  &      &      25      &  10    &    50\\
04118-3207 &       &  10-30  &    0-40  &    0-15     &16     &    0   &     40\\
04210-4042 &       &   0-5   &    0-5   &    0-5      & 3     &    0   &      5\\
04315-0840 &       &   0-15  &    0-10  &    0-5      & 6     &    0   &     15\\
04335-2514 &       &         &          &             &       &        &     \\
04370-2416 &       &   0-5   &    0-5   &    0-5      & 3     &    0   &      5\\
04461-0624$^{*}$ &       &   100   &    100   &    100      &100    &   100  &     100\\
04501-3304 &       &   0-5   &    0-5   &             & 3     &    0   &      5\\
04558-0751 &       &         &          &             &       &        &     \\
04569-0756 &       &         &          &             &       &        &     \\
04591-0419 &       &   0-0   &    0-10  &    0-5      & 3     &    0   &     10\\
04595-1813 &       &   0-5   &    0-10  &             & 4     &    0   &     10\\
05041-4938 &       &   0-5   &    0-5   &             & 3     &    0   &      5\\
05053-0805 &       &   0-5   &    0-0   &    0-5      & 2     &    0   &      5\\
05140-6213 &       &   0-10  &    0-10  &             & 5     &    0   &     10\\
05449-0651 &       &         &          &             &       &        &      \\
05562-6933 &       &   0-5   &    0-5   &    0-5      & 3     &    0   &      5\\
06295-1735 &       &   0-10  &    0-10  &    0-10     & 5     &    0   &     10\\
06592-6313 &       &   0-5   &    0-5   &    0-5      & 3     &    0   &      5\\
08175-1433 &       &         &          &             &       &        &       \\
08225-6936 &       &   0-5   &          &    0-5      & 3     &    0   &      5\\
08364-1430 &       &   0-5   &    0-5   &    0-10     & 3     &    0   &     10\\
08438-1510 &       &   0-10  &    0-40  &    0-15     &11     &    0   &     40\\
09006-6404 &       &   0-10  &    0-40  &    0-15     &11     &    0   &     40\\
09248-1918 &       &   0-5   &    0-10  &    0-10     & 4     &    0   &     10\\
09375-6951 &     1 &   0-10  &    0-10  &             & 5     &    0   &     10\\
09375-6951 &     2 &   0-5   &    0-5   &    0-5      & 3     &    0   &      5\\
10015-0614 &     1 &   0-5   &    0-5   &    0-5      & 3     &    0   &      5\\
10015-0614 &     2 &   0-5   &    0-10  &    0-5      & 3     &    0   &     10\\
10221-2317 &       &   0-20  &    0-20  &    0-20     &10     &    0   &     20\\
10409-4557 &       &   0-15  &    0-15  &             & 8     &    0   &     15\\
10484-0153 &       &   0-0   &    0-0   &    0-0      & 0     &    0   &      0\\
10567-4310 &       &         &          &             &       &        &       \\
11005-1601 &       &   0-5   &    0-5   &    0-5      & 3     &    0   &      5\\
11254-4120 &       &   0-5   &    0-5   &    0-10     & 3     &    0   &     10\\
11328-4844 &       &         &          &             &       &        &       \\
11409-1631 &       &   0-5   &    0-10  &    0-5      & 3     &    0   &     10\\
12042-3140 &     1 &   0-15  &          &    LINER    & 8     &    0   &     15\\
12042-3140 &     2 &   0-10  &          &             & 5     &    0   &     10\\
12112-4659 &       &   0-5   &    0-5   &    0-5      & 3     &    0   &      5\\
12115-4657 &       &   0-10  &    0-20  &    0-5      & 6     &    0   &     20\\
12120-1118 &       &   0-5   &    0-5   &    0-5      & 3     &    0   &      5\\
12171-1156 &       &   0-5   &    0-5   &             & 3     &    0   &      5\\
12286-2600 &       &   0-5   &    0-5   &    0-5      & 3     &    0   &      5\\
12329-3938$^{*}$ &       &  80-100 &   70-100 &    100      &92     &   70   &    100\\
12351-4015 &     1 &   0-5   &    0-5   &    0-5      & 3     &    0   &      5\\
12351-4015 &     3 &   0-5   &    0-10  &    LINER    & 4     &    0   &     10\\
12351-4015 &     2 &   0-5   &    0-5   &    0-5      & 3     &    0   &      5\\
12596-1529 &     1 &   0-10  &    0-5   &    0-5      & 3     &    0   &     10\\
12596-1529 &     2 &   0-5   &    0-10  &    0-5      & 3     &    0   &     10\\
12596-1529 &     3 &   0-0   &    0-20  &    0-10     & 8     &    0   &     20\\
13001-2339 &       &   0-10  &   10-30  &   LINER     &7      &   0    &    10\\
13035-4008$^{*}$ & &  50-60  &    100   &             &78     &   50   &    100\\
13097-1531 &       &   0-5   &    0-5   &    0-10     & 3     &    0   &      5\\
13135-2801 &       &   0-5   &    0-5   &             & 3     &    0   &      5\\
13192-5208 &       &   0-5   &    0-5   &             & 3     &    0   &      5\\
13197-1627$^{*}$ & &   100   &    100   &    100      &100    &  100   &   100\\
13229-2934$^{*}$ & &  30-40  &   25-50  &   10-20     &  31   &  10    &    50\\
14544-4255$^{*}$ & &  40-50  &   80-100 &   30-40     &  57   &  40    &   100\\
14566-1629$^{*}$ & &  30-50  &   40-50  &             &  43   &  30    &    50\\
15555-6610$^{*}$ & &   100   &    100   &    100      & 100   &  100   &    100\\
16153-7001 &       &         &          &             &       &        &      \\
16229-6640 &     1 &   0-5   &    0-5   &             & 3     &    0   &      5\\
16229-6640 &     2 &   0-5   &          &             & 3     &    0   &      5\\
17138-1017 &       &  10-10  &    0-10  &    0-5      & 4     &    0   &     10\\
17182-7353 &       &         &          &             &       &        &       \\
17260-7622 &       &   0-10  &    0-10  &             & 5     &    0   &     10\\
18093-5744 &       &   0-10  &    0-10  &    0-10     & 5     &    0   &     10\\
18293-3413 &       &         &          &             &       &        &      \\
18341-5732 &       &   0-10  &    0-10  &    0-10     & 5     &    0   &     10\\
18421-5049 &       &   0-20  &    0-20  &             &10     &    0   &      20\\
18429-6312$^{*}$ & &   100   &    100   &    100      & 100   &   100  &     100 \\
19543-3804$^{*}$ & &  80-100 &   90-100 &   100       &95     &   80   &      100\\
20305-0211 &     1 &   0-5   &    0-10  &    0-5      & 7     &    0   &     10\\
20305-0211 &     3 &   0-5   &    0-10  &    0-10     & 8     &    0   &     10\\
20309-1132 &       &   0-5   &    0-5   &             & 3     &    0   &      5\\
20486-4857 &       &         &          &             &       &        &      \\
21008-4347 &       &   0-5   &    0-5   &             & 3     &    0   &      5\\
21314-4102 &       &   0-5   &    0-10  &    L        &       &        &       \\
21330-3846 &     2 &   0-10  &    0-10  &    0-10     & 5     &    0   &     10\\
21330-3846 &     1 &   0-10  &    0-10  &             & 5     &    0   &     10\\
21453-3511$^{*}$ & &  40-45  &   35-50  &   40-50     &  43   &  35    &    50\\
22115-3013 &       &   0-5   &    0-5   &             & 3     &    0   &     10\\
22118-2742 &       &   0-10  &    0-10  &             & 5     &    0   &     10\\
22179-2455 &       &         &          &             &       &        &     \\
23394-0353 &       &   0-5   &    0-5   &    0-10     & 7     &    0   &     10\\
\enddata
\end{deluxetable}

\begin{deluxetable}{lcccccccccccccc}
\tabletypesize{\scriptsize}
\rotate
\tablewidth{0pc}
\tablecolumns{15}
\tablecaption{The multi-wavelength properties of the galaxies. }
\tablecomments{The 12, 25, 60, 100 $\mu$m fluxes are taken from the {\it IRAS Faint Source Catalogue}. The ATCA 4.8GHz total radio power ($P_{4.8}$) and the high resolution LBA flux ($S_{LBA}$) of the galaxies are taken from Paper I. P$_{FIR}$ is the FIR power of the galaxies. For a source with flux S$_{\nu}$ at a given frequency $\nu$ the power P$_{\nu}$ is defined as $4\pi D_{L} S_{\nu}$. $q$ is defined as Log($P_{FIR}/P_{4.8}$). Units are as follows: (2) kms$^{-1}$, (3) Mpc, (4-7) Jy, (10) $\times$ 10$^{24}$ W Hz$^{-1}$ (11) $\times$ 10$^{21}$ W Hz$^{-1}$ (13) mJy}
\tablehead{
\colhead{Name}&\colhead{Vel.}&\colhead{D}&\colhead{S$_{12}$}&\colhead{S$_{25}$}&\colhead{S$_{60}$}&\colhead{S$_{100}$}&\colhead{$\frac{S_{60}}{S_{25}}$}&\colhead{Log($\frac{L_{IR}}{L\sun}$)}&\colhead{P$_{FIR}$}&\colhead{P$_{4.8}$}&\colhead{q} &\colhead{S$_{LBA}$}&\colhead{V95 Class} & \colhead{K01 Class}\\
\colhead{(1)}& \colhead{(2)}& \colhead{(3)}& \colhead{(4)}& \colhead{(5)}& \colhead{(6)}& \colhead{(7)}& \colhead{(8)}& \colhead{(9)}& \colhead{(10)}& \colhead{(11)}& \colhead{(12)}& \colhead{(13)}& \colhead{(14)}& \colhead{(15)}\\
}
\startdata
00085-1223 & 5821 & 77.99 &   0.40 & 2.37 & 16.60 & 17.20 &  7.00  &  11.42 & 14.67  & 2.15 &  2.83 & $<$2.3 &AMB    & AMB\\
00344-3349 & 6156 & 82.50 &   0.42 & 2.49 &  6.48 &  5.01 &  2.60  &  11.18 &  5.94  & 1.24 &  2.68 & $<$1.5 &HII/HII&HII/HII\\
00402-2350 & 6647 & 89.12 &   0.33 & 1.08 & 10.00 & 18.30 &  9.26  &  11.37 & 14.07  & 2.17 &  2.81 & $<$2.6 &L      &   HII\\
01053-1746 & 6016 & 80.62 &   0.68 & 3.57 & 22.60 & 30.40 &  6.33  &  11.62 & 23.17  & 5.80 &  2.60 & $<$1.7 &HII    & HII\\
01159-4443 & 6701 & 89.85 &   0.30 & 1.86 &  7.84 & 11.70 &  4.22  &  11.31 & 10.36  & 2.08 &  2.70 & $<$1.4 &HII/B  & HII/AMB\\
01165-1719 & 5977 & 80.09 &   0.19 & 0.63 &  4.06 &  6.99 &  6.44  &  10.92 &  4.50  & 0.87 &  2.72 & $<$1.7 &HII    & HII\\
01326-3623 & 4827 & 64.62 &   0.26 & 0.68 &  7.29 & 15.80 & 10.77  &  10.98 &  5.81  & 1.53 &  2.58 & $<$1.9 &N      &   N\\
01341-3734 & 5180 & 69.37 &   0.27 & 0.69 &  5.39 & 12.80 &  7.87  &  10.95 &  5.16  & 0.60 &  2.94 & $<$1.8 &HII/HII& HII/HII\\
01384-7515 & 3966 & 53.05 &   0.33 & 0.81 &  7.28 & 13.80 &  8.98  &  10.81 &  3.69  & 0.52 &  2.85 & $<$1.1 &HII    & HII\\
02015-2333 & 4934 & 66.06 &   0.20 & 0.76 &  4.02 &  7.32 &  5.31  &  10.77 &  3.10  & 0.51 &  2.78 & $<$2.7 &HII    & HII\\
02069-1022 & 4152 & 55.55 &   0.35 & 0.45 &  5.31 & 17.70 & 11.77  &  10.82 &  3.89  & 0.46 &  2.93 & $<$1.9 &B      & HII\\
02072-1025 & 3847 & 51.46 &   0.45 & 2.48 & 10.60 & 23.60 &  4.27  &  11.01 &  5.42  & 0.62 &  2.94 & $<$1.6 &B      &   HII\\
02140-1134 & 4009 & 53.63 &   0.44 & 0.75 &  5.97 & 11.80 &  7.98  &  10.77 &  3.14  & 0.55 &  2.76 & $<$2.2 &HII    & HII\\
02281-0309 & 5738 & 76.87 &   0.47 & 0.53 &  5.25 & 15.60 & 10.00  &  11.10 &  6.92  & 1.44 &  2.68 & $<$1.9 &N      &   N\\
02433-1534 & 4101 & 54.87 &   0.32 & 0.68 &  7.12 & 13.80 & 10.47  &  10.82 &  3.89  & 0.73 &  2.73 & $<$1.8 &AMB    & HII\\
02436-5556 & 5507 & 73.76 &   0.29 & 0.44 &  4.66 & 11.40 & 10.66  &  10.95 &  5.12  & 0.72 &  2.85 & $<$1.8 &N      &   N\\
02476-3858 & 5008 & 67.05 &   0.08 & 0.33 &  4.47 &  6.68 & 13.59  &  10.72 &  3.29  & 0.34 &  2.99 & $<$1.8 &N      &   N\\
03022-1232 & 4282 & 57.30 &   0.27 & 0.99 &  7.76 & 10.60 &  7.85  &  10.85 &  4.04  & 0.55 &  2.87 & $<$1.6 &Sy2    &  Sy2\\
04118-3207 & 3570 & 47.74 &   0.53 & 2.13 & 14.20 & 21.60 &  6.67  &  10.98 &  5.33  & 0.71 &  2.88 & $<$2.2 &B      &   B(AGN)\\
04210-4042 & 6012 & 80.56 &   0.23 & 0.78 &  8.01 & 16.40 & 10.32  &  11.20 &  9.67  & 1.42 &  2.83 & $<$1.9 &HII    & HII\\
04315-0840 & 4744 & 63.50 &   1.44 & 7.29 & 32.30 & 32.70 &  4.43  &  11.59 & 18.80  & 3.06 &  2.79 & $<$2.8 &HII    & HII\\
04335-2514 & 4843 & 64.83 &   0.17 & 0.43 &  4.99 &  9.75 & 11.63  &  10.80 &  3.82  & 0.80 &  2.68 & $<$2.5 &N      &   N\\
04370-2416 & 4422 & 59.18 &   0.34 & 0.57 &  6.00 & 11.00 & 10.51  &  10.82 &  3.73  & 0.66 &  2.75 & $<$1.5 &HII    & HII\\
04461-0624 & 4578 & 61.27 &   0.43 & 0.68 &  5.95 & 14.70 &  8.79  &  10.91 &  4.53  & 1.06 &  2.63 & $<$1.8 &Sy2    & Sy2\\
04502-3304 & 5622 & 75.31 &   0.20 & 1.18 &  8.39 &  9.40 &  7.11  &  11.10 &  7.08  & 0.61 &  3.06 & $<$1.6 &AMB    & HII\\
04558-0751 & 3773 & 50.47 &   0.20 & 0.60 &  5.94 & 10.10 &  9.85  &  10.64 &  2.60  & 0.41 &  2.80 & $<$1.4 &N      &  N\\
04569-0756 & 4242 & 56.76 &   0.29 & 0.71 &  6.78 & 13.60 &  9.55  &  10.84 &  4.03  & 0.55 &  2.87 & $<$2.7 &N      &   N\\
04591-0419 & 4068 & 54.42 &   0.11 & 0.58 &  3.92 &  5.84 &  6.77  &  10.52 &  1.90  & 0.34 &  2.74 & $<$1.5 &HII    & HII\\
04595-1813 & 3978 & 53.22 &   0.19 & 0.34 &  4.05 &  8.78 & 12.05  &  10.57 &  2.19  & 0.40 &  2.74 & $<$2.3 &HII    & HII\\
05041-4938 & 4145 & 55.46 &   0.25 & 0.29 &  3.07 &  7.88 & 10.70  &  10.54 &  1.95  & 0.10 &  3.28 & $<$1.6 &HII    & HII\\
05053-0805 & 4478 & 59.93 &   0.33 & 1.33 &  8.26 & 13.30 &  6.21  &  10.96 &  5.00  & 0.46 &  3.03 & NObs   &HII    & HII\\
05140-6213 & 4966 & 66.49 &   0.18 & 0.33 &  2.78 & 10.20 &  8.53  &  10.72 &  3.09  & 0.20 &  3.19 & $<$1.7 &HII    & HII\\
05449-0651 & 6467 & 86.69 &   0.25 & 0.25 &  0.85 &  6.02 &  3.40  &  10.73 &  2.48  & ND   &  ...  & $<$1.7 &N      &   N\\
05562-6933 & 4440 & 59.42 &   0.27 & 0.58 &  4.48 & 15.10 &  7.74  &  10.81 &  3.78  & 0.45 &  2.92 & $<$1.7 &HII    & HII\\
06295-1735 & 6339 & 84.97 &   0.11 & 0.56 &  6.36 &  9.12 & 11.44  &  11.08 &  7.41  & 0.78 &  2.98 & $<$1.5 &HII    & HII\\
06592-6313 & 6882 & 92.29 &   0.16 & 0.80 &  5.74 &  7.56 &  7.20  &  11.13 &  7.66  & 0.88 &  2.94 & $<$2.6 &AMB    & HII\\
08175-1433 & 5732 & 76.79 &   0.21 & 0.56 &  4.63 &  9.02 &  8.33  &  10.94 &  4.97  & 1.13 &  2.64 & $<$1.3 &N      &   N\\
08225-6936 & 3924 & 52.49 &   0.20 & 0.67 &  4.72 &  6.41 &  7.00  &  10.58 &  2.06  & 0.29 &  2.85 & $<$1.7 &HII    & HII\\
08364-1430 & 4184 & 55.98 &   0.46 & 0.46 &  4.23 &  7.40 &  9.18  &  10.67 &  2.31  & 0.46 &  2.70 & $<$1.3 &HII    & HII\\
08438-1510 & 5423 & 72.63 &   0.13 & 0.57 &  4.11 &  5.63 &  7.17  &  10.79 &  3.44  & 0.57 &  2.78 & $<$2.1 &L      & B(HII)\\
09006-6404 & 6636 & 88.97 &   0.15 & 0.48 &  3.87 &  5.88 &  8.00  &  10.95 &  5.05  & 1.09 &  2.67 & $<$1.6 &HII    & HII\\
09248-1918 & 4888 & 65.44 &   0.15 & 0.56 &  4.16 &  7.61 &  7.50  &  10.74 &  3.16  & 0.51 &  2.80 & $<$2.5 &HII    & HII\\
09375-6951 & 6066 & 81.29 &   0.20 & 0.41 &  4.34 &  9.68 & 10.48  &  10.97 &  5.54  & 1.38 &  2.61 &  3.6   &B/HII  &  HII/HII\\
10015-0614 & 5034 & 67.40 &   0.59 & 1.04 & 10.70 & 19.20 & 10.29  &  11.18 &  8.54  & 2.06 &  2.62 & $<$1.6 &HII/HII& HII/HII\\
10221-2317 & 3662 & 48.98 &   0.30 & 1.20 & 11.20 & 14.80 &  9.33  &  10.85 &  4.21  & 0.65 &  2.81 & $<$2.7 &B      &   HII\\
10409-4557 & 7000 & 93.88 &   0.37 & 0.64 &  5.28 & 12.60 &  8.26  &  11.22 &  9.29  & 1.42 &  2.82 & $<$1.7 &L      &   HII\\
10484-0153 & 4464 & 59.74 &   0.28 & 0.54 &  4.51 &  9.83 &  8.29  &  10.74 &  3.08  & ND   &  ...  & $<$1.3 &HII    & HII\\
10567-4310 & 5156 & 69.04 &   0.35 & 0.80 &  5.76 & 10.40 &  7.17  &  10.95 &  4.84  & 0.64 &  2.88 & $<$1.6 &N      &   N\\
11005-1601 & 3877 & 51.86 &   0.41 & 0.89 &  6.75 & 14.00 &  7.55  &  10.79 &  3.40  & 0.55 &  2.79 & $<$2.6 &HII    & HII\\
11254-4120 & 4902 & 65.63 &   0.20 & 0.81 &  8.06 &  8.41 &  9.94  &  10.93 &  5.05  & 0.53 &  2.98 & $<$1.3 &AMB    & HII\\
11328-4844 & 5624 & 75.34 &   0.20 & 0.43 &  4.01 &  7.11 &  9.33  &  10.84 &  3.98  & 0.90 &  2.65 & $<$1.6 &N      &   N\\
11409-1631 & 3660 & 48.95 &   0.21 & 0.63 &  4.20 &  7.01 &  6.71  &  10.50 &  1.72  & 0.24 &  2.86 & $<$1.6 &HII    & HII\\
12042-3140 & 6818 & 91.42 &   0.25 & 0.68 &  7.56 & 11.40 & 11.13  &  11.24 & 10.38  & 1.91 &  2.74 & $<$1.7 &B/HII  & AMB/HII\\
12112-4659 & 5493 & 73.58 &   0.19 & 0.66 &  4.71 &  6.81 &  7.15  &  10.87 &  4.12  & 0.46 &  2.95 & $<$1.5 &HII    & HII\\
12115-4657 & 5543 & 74.25 &   0.32 & 0.66 &  5.23 &  9.00 &  7.90  &  10.96 &  4.98  & 1.19 &  2.62 & $<$1.7 &HII    & HII\\
12120-1118 & 5406 & 72.41 &   0.25 & 0.38 &  4.01 &  6.70 & 10.69  &  10.81 &  3.59  & 0.73 &  2.69 & $<$1.6 &HII    & HII\\
12171-1156 & 4234 & 56.65 &   0.22 & 0.51 &  4.54 &  6.65 &  8.85  &  10.63 &  2.37  & 0.20 &  3.08 & $<$1.9 &HII    & HII\\
12286-2600 & 5970 & 80.00 &   0.17 & 0.59 &  3.95 &  6.49 &  6.73  &  10.89 &  4.29  & 0.51 &  2.93 & $<$1.4 &HII    & HII\\
12329-3938 & 3523 & 47.11 &   0.46 & 1.39 &  4.31 &  5.40 &  3.10  &  10.56 &  1.47  & 0.47 &  2.50 &  1.6   &Sy2     &  Sy2\\
12351-4015 & 5250 & 70.31 &   0.45 & 0.84 &  7.11 & 16.00 &  8.47  &  11.09 &  6.82  & 0.95 &  2.86 & $<$2.0 &HII/AMB/AMB&HII/HII/AMB\\
12596-1529 & 4773 & 63.89 &   0.34 & 1.36 &  7.38 &  9.09 &  5.43  &  10.95 &  4.61  & 0.60 &  2.88 & $<$1.6 &HII/HII/HII&HII/HII/HII\\
13001-2339 & 6446 & 86.41 &   0.14 & 0.85 & 13.70 & 15.30 & 16.19  &  11.37 & 15.20  & 3.28 &  2.67 & $<$1.4 &L      &   AMB\\
13035-4008 & 4475 & 59.89 &   0.68 & 1.22 &  5.66 &  8.79 &  4.64  &  10.89 &  3.37  & 0.54 &  2.79 & $<$1.3 &Sy1    & Sy1\\
13097-1531 & 6400 & 85.79 &   0.34 & 0.96 & 11.10 & 20.90 & 11.51  &  11.38 & 14.65  & 2.58 &  2.75 & 2.5    &B      &   HII\\
13135-2801 & 4426 & 59.23 &   0.18 & 0.71 &  4.72 &  8.91 &  6.63  &  10.72 &  2.97  & 0.37 &  2.91 & 2.2    &HII    & HII\\
13193-5208 & 5090 & 68.15 &   0.20 & 0.67 &  4.41 &  6.05 &  6.54  &  10.78 &  3.25  & 0.28 &  3.06 & $<$2.0 &HII    & HII\\
13197-1627 & 5152 & 68.99 &   0.88 & 2.86 &  5.89 &  5.48 &  2.06  &  11.09 &  3.95  & 5.57 &  1.85 & 8.9    &Sy1.8  & Sy1.8\\
13229-2934 & 4112 & 55.01 &   0.64 & 2.40 & 16.90 & 28.60 &  7.04  &  11.19 &  8.78  & 2.50 &  2.55 & $<$1.5 &Sy2    & Sy2\\
14544-4255 & 4875 & 65.26 &   0.35 & 1.31 &  7.61 & 12.80 &  5.81  &  11.01 &  5.55  & 2.19 &  2.40 & 3.8 &Sy2    & Sy2\\
14566-1629 & 3521 & 47.08 &   0.19 & 0.59 &  6.21 &  8.96 & 10.49  &  10.57 &  2.23  &11.81 &  1.28 & 266.1  &Sy2    & Sy2\\
15555-6610 & 3638 & 48.65 &   0.21 & 0.30 &  2.41 & 12.40 &  8.09  &  10.48 &  1.77  & ND   &   ... & $<$1.5 &Sy2    & Sy2\\
16153-7001 & 3542 & 47.37 &   0.48 & 0.68 &  6.85 & 17.90 & 10.07  &  10.75 &  3.21  & 0.37 &  2.93 & $<$1.3 &N      &   N\\
16229-6640 & 6535 & 87.61 &   0.19 & 0.50 &  4.42 &  8.99 &  8.86  &  11.03 &  6.29  & 0.90 &  2.84 & $<$1.4 &AMB/HII& HII/HII\\
17138-1017 & 5261 & 70.45 &   0.62 & 2.07 & 15.20 & 19.00 &  7.34  &  11.33 & 11.61  & 1.51 &  2.89 & $<$1.5 &HII    & HII\\
17182-7353 & 4788 & 64.10 &   0.16 & 0.27 &  3.58 &  9.30 & 13.16  &  10.70 &  3.06  & 0.60 &  2.71 & $<$1.7 &N      &   N\\
17260-7622 & 5507 & 73.76 &   0.17 & 0.43 &  4.07 &  5.36 &  9.38  &  10.79 &  3.47  & 0.91 &  2.58 & $<$1.6 &HII    & HII\\
18093-5744 & 5200 & 69.63 &   0.65 & 2.38 & 15.20 & 25.10 &  6.39  &  11.36 & 12.53  & 1.81 &  2.84 & $<$1.4 &HII    & HII\\
18293-3413 & 5449 & 72.98 &   1.12 & 3.76 & 34.20 & 49.70 &  9.10  &  11.71 & 29.52  & 5.06 &  2.77 & $<$2.1 &N      &   N\\
18341-5732 & 4601 & 61.58 &   0.37 & 1.27 & 14.40 & 25.20 & 11.34  &  11.19 &  9.50  & 1.08 &  2.94 & $<$1.8 &AMB    & HII\\
18421-5049 & 5266 & 70.52 &   0.15 & 0.52 &  5.27 &  7.97 & 10.21  &  10.86 &  4.31  & 0.58 &  2.87 & $<$1.5 &B      & HII\\
18429-6312 & 4367 & 58.44 &   0.23 & 0.76 &  4.38 &  8.15 &  5.75  &  10.70 &  2.67  & 0.32 &  2.92 & $<$1.8 &Sy2     & Sy2\\
19543-3804 & 5713 & 76.54 &   0.37 & 1.17 &  6.05 &  9.14 &  5.17  &  11.06 &  5.83  & 4.32 &  2.13 & 23.4   &Sy2     & Sy2\\
20305-0211 & 5970 & 80.00 &   0.36 & 0.51 &  5.59 & 14.30 & 10.92  &  11.11 &  7.39  & 3.01 &  2.39 & $<$1.7 &HII/HII& HII/HII\\
20309-1132 & 3549 & 47.46 &   0.22 & 0.45 &  4.30 &  7.31 &  9.56  &  10.47 &  1.67  & 0.24 &  2.84 & $<$1.7 &HII    & HII\\
20486-4857 & 5288 & 70.82 &   0.31 & 0.44 &  4.51 &  9.04 & 10.20  &  10.88 &  4.17  & 0.66 &  2.80 & $<$1.7 &N      &   N\\
21008-4347 & 5208 & 69.74 &   0.29 & 0.78 &  6.93 & 12.60 &  8.94  &  11.01 &  5.96  & 0.88 &  2.83 & $<$1.7 &HII    & HII\\
21314-4102 & 5161 & 69.11 &   0.26 & 0.46 &  3.95 &  9.25 &  8.55  &  10.83 &  3.73  & 0.46 &  2.91 & $<$1.4 &L      & AMB\\
21330-3846 & 5714 & 76.55 &   0.28 & 0.88 &  5.86 &  8.91 &  6.66  &  11.02 &  5.66  & 0.72 &  2.90 & $<$1.7 &L/HII  & HII/HII\\
21453-3511 & 4842 & 64.82 &   0.59 & 2.12 & 16.50 & 25.60 &  7.78  &  11.31 & 11.51  & 3.25 &  2.55 &  4.7   &Sy2     &  Sy2\\
22115-3013 & 4291 & 57.42 &   0.15 & 0.84 &  4.15 &  4.34 &  4.96  &  10.59 &  1.99  & 0.28 &  2.85 & $<$1.7 &HII    & HII\\
22118-2742 & 5247 & 70.27 &   0.41 & 0.58 &  5.46 & 12.10 &  9.41  &  10.98 &  5.20  & 0.96 &  2.73 & $<$2.6 &HII    & HII\\
22179-2455 & 4688 & 62.75 &   0.24 & 0.43 &  3.97 &  7.02 &  9.19  &  10.69 &  2.73  & 0.44 &  2.79 & $<$2.5 &N      &   N\\
23394-0353 & 6966 & 93.42 &   0.98 & 0.45 &  0.69 &  1.61 &  1.55  &  10.96 &  1.19  & 1.82 &  1.82 & $<$1.5 &HII    & HII\\
\enddata
\end{deluxetable}

\begin{deluxetable}{lcccccc}
\tabletypesize{\footnotesize}
\footnotesize
\tablewidth{0pc}
\tablecolumns{7}
\tablecaption{Table summarizing the distribution of galaxy spectral classifications (based on the K01 system) with luminosity. The distribution for galaxies classified with the V95 system is shown in brackets.}
%\tablecomments{}
\tablehead{ 
	\colhead{log(L$_{\rm IR}/$L$\sun$)} & \colhead{HII} & \colhead{Sy} & \colhead{LINERs} & \colhead{Borderline} &\colhead{Ambiguous}&\colhead{Not classified} }
\startdata
$<$ 10.75      & 13 (13) & 4 (4) & 0 (0) & 0 (0) &0 (0) &5 (5) \\   
10.75-11.00    & 26 (20) & 3 (3) & 0 (2) & 2 (5) &1 (2) &8 (8) \\
11.00-11.25    & 14 ( 7) & 4 (4) & 0 (2) & 0 (2) &1 (4) &1 (1) \\ 
$>$ 11.25      &  7 ( 5) & 1 (1) & 0 (2) & 0 (1) &2 (1) &1 (1) \\
\enddata 
\end{deluxetable}

\begin{deluxetable}{lcccc}
\tabletypesize{\footnotesize}
\footnotesize
\tablewidth{0pc}
\tablecolumns{5}
\tablecaption{Table showing the mean, median and standard deviation of the value $q$ for the COLA sample as a function of the (K01) spectral classification.}
\tablecomments{$^{1}$ One starburst galaxy (IRAS 05035-0805) was not observed with the LBA and is therefore excluded from the sample of HII galaxies without cores.}
\tablehead{ 
	\colhead{Galaxy spectral type} & \colhead{No.} & \colhead{Mean
	q} & \colhead{Median q} & \colhead{$\sigma$}
 }
\startdata
All galaxies       & 90 & 2.76 & 2.80 & 0.27 \\   
All Seyferts       & 11 & 2.54 & 2.41 & 0.48 \\
{\it Sy w. cores}  &  6 & 2.12 & 2.27 & 0.49 \\ 
{\it Sy w/o cores} &  5 & 2.75 & 2.79 & 0.16 \\
All HII            & 59 & 2.81 & 2.83 & 0.20 \\
{\it HII w. cores} &  3 & 2.75 & 2.75 & 0.15 \\ 
{\it HII w/o cores}& 55$^{1}$ & 2.81 & 2.84 & 0.21 \\
Ambiguous          &  4 & 2.79 & 2.78 & 0.11 \\
Borderline         &  2 & 2.83 & 2.83 & 0.07 \\ 
Not classified     & 15 & 2.77 & 2.79 & 0.12 \\
\enddata 
\end{deluxetable}

\pagebreak

\clearpage
\begin{figure}
\figurenum{1}
\epsscale {0.5}
\plotone{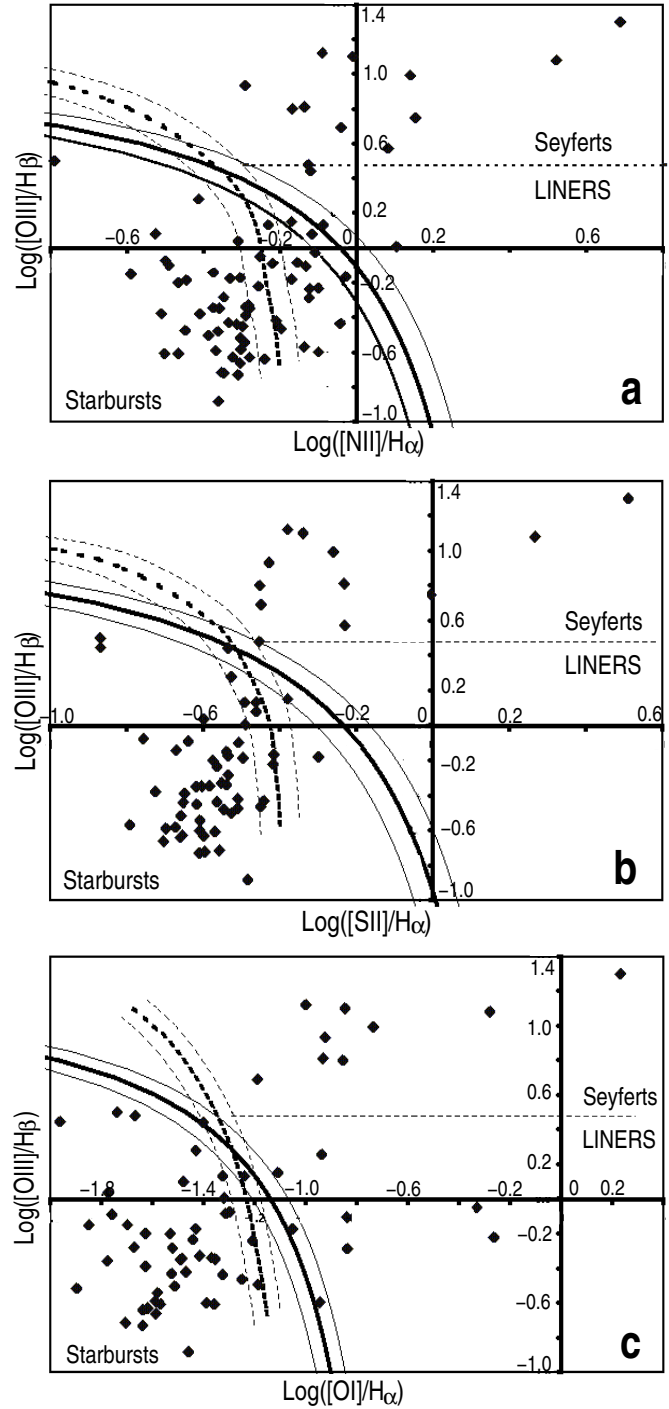}
\caption[]{Diagnostic diagrams for emission line ratios. The dotted
lines indicate the HII/AGN partition ($\pm$0.5dex) for the V95
classification system and the solid lines show the theoretical HII/AGN
partition ($\pm$0.5dex) from the K01 system.}
\end{figure} 
\clearpage
\begin{figure}
\figurenum{2}
\epsscale{1}
\plotone{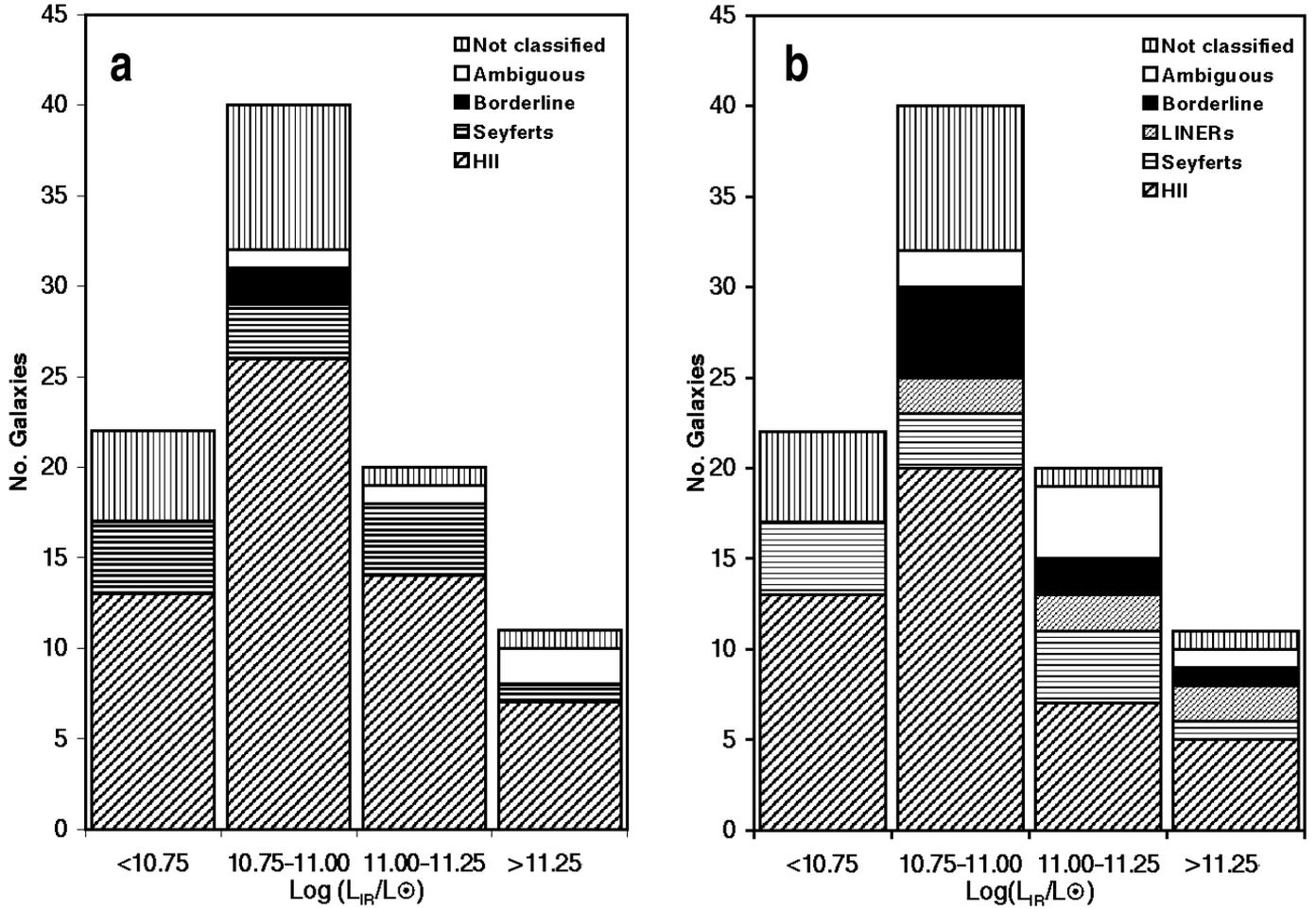}
\caption[]{
a)The distribution of the classification of the galaxies of our sample
as a function of L$_{\rm IR}$ using the K01 scheme. b) Same as in a)
 but using the V95 scheme. Note that even though no significant trend
 with luminosity was seen in the Seyferts, the proportion of LINERS,
 borderline and ambiguous galaxies rises when one uses the V95 system.}
\end{figure} 

\clearpage
\begin{figure}
\figurenum{3}
\epsscale {0.5}
\plotone{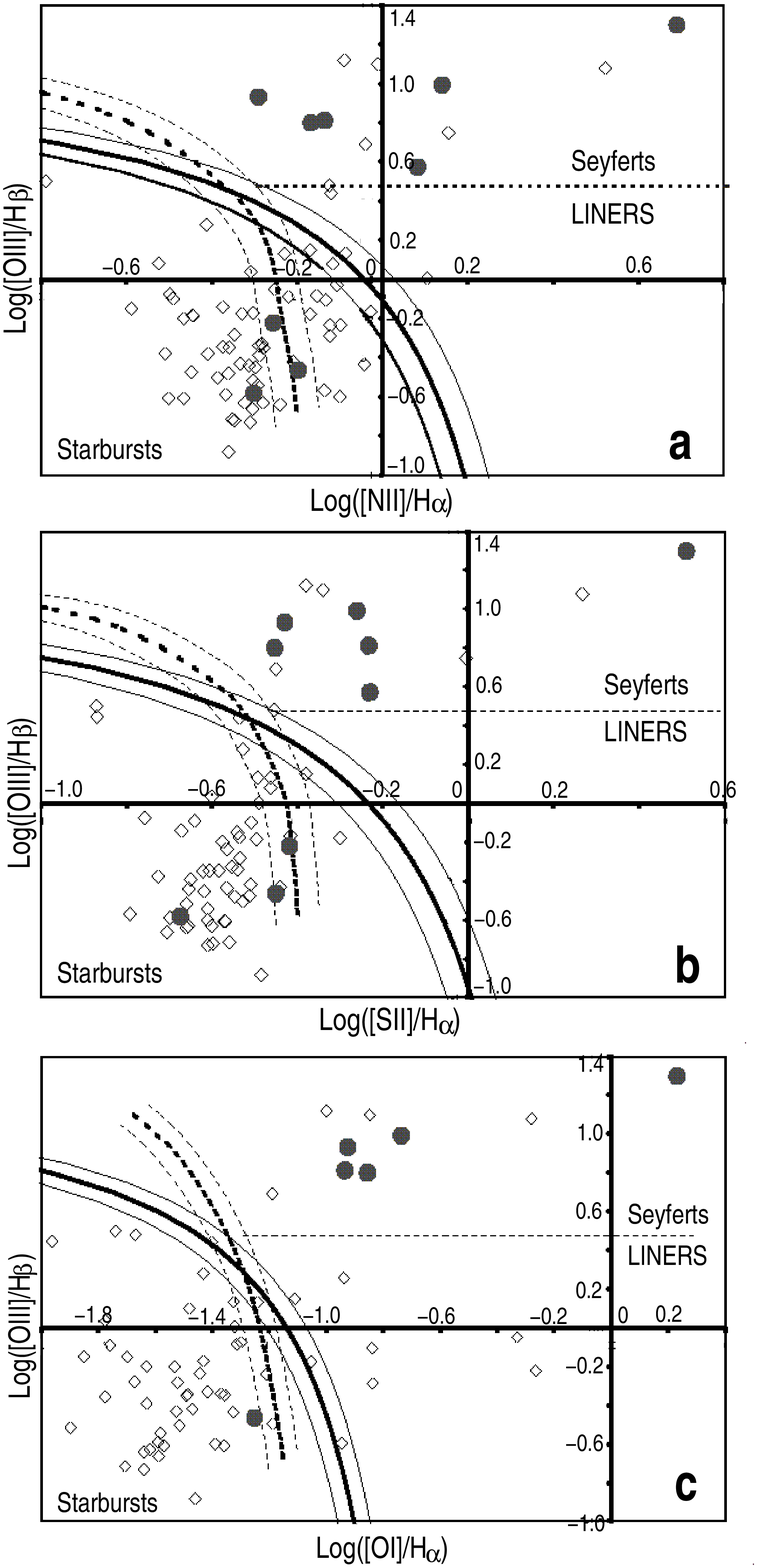}
\caption[]{The same diagnostic diagrams presented in Fig. 1 with the 
galaxies in which compact radio cores were detected marked with filled
symbols.}
\end{figure} 

\clearpage
\begin{figure}
\figurenum{4}
\epsscale{0.5}
\plotone{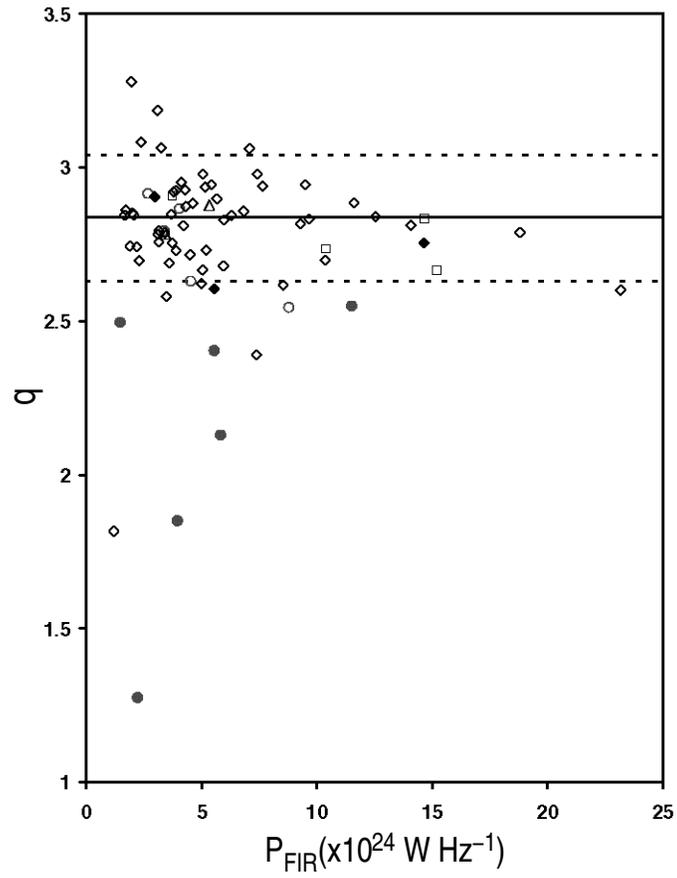}
\caption[]{Plot of $q$ versus the galaxy FIR power. Galaxies classified
 as starburst galaxies are represented by
 diamonds, Seyferts by circles, borderline galaxies by triangles and
 ambiguous classifications by squares. Filled symbols
 indicate those galaxies in which a compact radio core were detected. A
 solid line indicates is shown at $q$=2.84, the median $q$ for the
 starburst galaxies without cores, with dotted lines at $\pm 1\sigma$.}
\end{figure} 
\clearpage
\appendix
\section{ Notes on individual galaxies}
\subsection{Seyfert glaxies in which compact radio cores were detected}  
\begin{itemize} 

\item {\bf 12329-3938 (NGC 4507)} (Sy2; $q$=2.5) is a barred spiral galaxy \citep[SAB(s)ab;][]{Sandage79}. It is a known Sy2 galaxy \citep{Degrijp87,Vaceli97}. A tentative detection of a $<0.05''$ core at 1.6\,mJy ($10^{3.4}$ L$\sun$) was reported in Paper I, making this source the lowest luminosity core detection in the sample.

\item {\bf IRAS 13197-1627} (Sy1.8; $q$=1.85) is a known Seyfert galaxy \citep{Degrijp85} 
and is classified as a Seyfert 1.8 by \citet{Aguero94}. The narrow emission lines in this object are unusually broad (FWHM of [OIII]$\lambda $5007 $\sim$1000kms$^{-1}$, FWHM of [NII]$\lambda $5007 $\sim$700kms$^{-1}$) but symmetrical. Evidence of a broad H$\alpha$ line is seen in our spectra but the H$\beta$ line has a similar width to the [OIII] lines.  It has a
tentative morphological classification in in the {\it NASA Extra-galactic Database} (NED) as a barred spiral galaxy. A compact ($<$ 0.05'') core of 8.9\,mJy was reported in Paper I and VLA
observations at 4.8GHz by \citet{Kinney00} have revealed a
linear structure extending about 278pc from the core, believed to be a
synchrotron jet.  IRAS 13197-1627 exhibits one of the lowest values of $q$ in our sample. This radio excess remains after subtraction
of the compact core emission and is probably due, at least in part, to
emission from the $\sim$ 300~pc-scale radio jet.

\item {\bf IRAS 14544-4255 (IC4518A)} (Sy2; $q$=2.4) is one of a pair of strongly interacting galaxies \citep{Condon96} and has not been previously identified as a Seyfert galaxy. This source exhibits a large radio excess and the 3.77\,mJy (10$^{4.1}$L$\sun$) compact radio core (Paper I) contributes only 3.5\% of the total radio emission.
 
\item {\bf IRAS 14566-1629 (NGC 5793)} (Sy2;$q$=1.28) has an Sb classification (NED)
and is one of a pair of galaxies. VLBI observations of this source
have detected a compact nucleus ($0.016''$) which is extended
parallel to the minor axis of the galaxy \citep{Gardner92}. There is
also some evidence for a water maser with a diameter larger than 20pc
\citep{Hagiwara97}. It is a known Seyfert 2 \citep{Baan98} and exhibits both the largest radio excess and the brightest radio core (10$^{5.6}$L$\sun$) of the galaxies in our sample (Paper I), accounting for 30\% of the observed radio flux. This source exhibits the coolest FIR colours of the Seyferts in our sample ($S_{60}/ S_{25}$=10.5) presumably indicating that star formation dominates the IR spectrum but not at optical or radio wavelengths.

\item {\bf IRAS 19543-3804} (Sy2; $q$=2.13) is classified as a spiral galaxy (Sb or
IrS) in a cluster. A compact ($<$0.05'') core was detected with a flux of 23.4\,mJy ($10^{4.9}{\rm L}_{\sun}$; Paper I) . This core is one of the
brightest we detect and accounts for 32\% of the total radio
luminosity at 2.3 GHz. This source exhibits a large radio excess and subtraction of the compact core
reduces the radio excess to $\sim$ 40\%.

\item {\bf IRAS 21453-3511 (NGC7130, IC5135)} (Sy2; $q$=2.55) is a known AGN,
identified by \citet{Phillips83} as a Seyfert 2 active nucleus
surrounded by a starburst ring. Observations by \citet{Thuan84}
further showed that the starburst dominates the galaxy emission,
contributing 75\% of the emission in the UV. Its optical emission line
ratios have confirmed its AGN classification but it falls on the
borderline between the Sy2 and LINER classes with \citet{Veilleux95}
classifying it as a LINER while both \citet{Heisler97} and \citet{Vaceli97}
classify it as a Seyfert 2. We also classify it as a Seyfert 2 and it
is possible that the differences between the classifications reflect
differing contributions from the starburst emission in the slit.

A 4.7\,mJy ($10^{4.1}$ ${\rm L}_{\sun}$) core was detected in IRAS
21453-3511 (Paper I) and the 4.8 GHz
luminosity of the galaxy is nearly twice that predicted from the FIR
luminosity. \citet{Heisler97} measured a compact core flux at 2.3 GHz of 14\,mJy
using the PTI baseline with a resolution of $0.1''$, twice that of
ours, which implies that the source has compact radio structures
between 25-50pc in size which we have resolved out. These structures
could be associated with a radio jet. Alternatively, the difference
between the PTI flux and our measurements could be due to intrinsic
variability of the compact core.
\end{itemize}

\subsection{Seyfert galaxies in which compact cores were not detected}
\begin{itemize}
\item {\bf IRAS 03022-1232 (NGC 1204)} (Sy2; $q$=2.87) was classified as a LINER galaxy by V95 with log(OIII/H$\beta$)=0.47. We measure log(OIII/H$\beta$)=0.48 classifying it as a Sy2, but in any case it clearly lies on the border between the LINER and Seyfert classes in the V95 system. It has an SO/a morphology (NED). No compact radio core was detected ($<$ 1.8mJy; Paper I), indicating that any radio core must have a luminosity $<$ 10$^{3.6}$L$\sun$. This source does not exhibit a radio excess indicating that the radio continuum is probably dominated by star formation. 

\item {\bf IRAS 04461-0624 (NGC1667)} (Sy2; $q$=2.63) is a known Seyfert galaxy \citep{Phillips83}. Although it appears that H$\alpha$ is slightly broadened in our spectrum, we find that both the [NII] emission lines are broadened in a similar fashion (this was also seen by \citet{Ho97a}) and we therefore conclude that no broad emission line component is visible. Spectropolarimetric observations of this source \citep{Tran01} have not revealed a hidden broad line region in this object. \citet{Thean00} measure a compact core ($<0.3''$) with a 4.8GHz flux of  1.5 mJy but we detected no compact radio ($<0.05''$) core ($<$ 1.6mJy; Paper I) indicating that any radio core must have a luminosity $<$ 10$^{3.7}$L$\sun$. Since one would expect it to be brighter at 2.3GHz than at 4.8GHz (assuming a spectral index of 0.7), our non-detection of a core implies that the core is either larger than $<0.05''$ and hence much of the flux is resolved out or has a flat spectral index. This source is one of the two Seyfert galaxies with cool FIR colours ($S_{60}/S_{25}$=8.8)
\item {\bf IRAS 13035-4008 (ESO 323-G77)} (Sy1; $q$=2.79) is a known Seyfert 1 galaxy \citep{Fairall86} and is classified in NED as a barred spiral galaxy.  Broad emission lines are clearly visible in both H$\alpha$ and H$\beta$. No compact radio core was detected ($<$ 1.3 mJy; Paper I ) indicating that any radio core must have a luminosity $<$ 10$^{3.5}$L$\sun$. This source does not exhibit a radio excess indicating that the radio continuum is probably dominated by star formation.
 
\item {\bf IRAS 13229-2934 (NGC5135)} (Sy2; $q$=2.55) is a known Sy2 galaxy \citep{Phillips83} and has barred spiral morphology (NED). No compact radio ($<0.05''$) core was detected ($<$ 1.5 mJy; Paper I) indicating that any radio core must have a luminosity $<$ 10$^{3.5}$L$\sun$. This source was also observed by \citet{Thean00} who failed to detect compact radio emission from the nucleus. IRAS 13229-2934 does exhibit a similar radio excess to IRAS 21453-3511 which may be due to the extended structure and radio lobes detected by \cite{Bransford98} and \cite{Thean00}.   

\item {\bf IRAS 18429-6312} (Sy2; $q$=2.92) is a known Sy2 galaxy \citep{Fairall79} and has an SB morphology (NED). No compact radio core was detected ($<$ 1.8 mJy ) indicating that any radio core must have a luminosity $<$ 10$^{3.7}$L$\sun$.  This source exhibits a radio {\it deficit} indicating that the radio continuum is dominated by star formation.
\end{itemize} 

\subsection{Non-Seyfert galaxies in which compact cores were detected}

\begin{itemize}
\item {\bf IRAS 09375-6951} (HII; q=2.61) is classified in the NASA Extragalactic
Database (NED) as an irregular spiral galaxy. We classify this object as a starburst using the K01 scheme and a borderline or composite object \citep[e.g.][]{Hill99,Barth00}.  A compact ($<0.05''$) core was detected
with a flux of 3.6\,mJy ($10^{4.25}{\rm L}_{\sun}$; Paper I) which
contributes 13\% of the total radio emission. IRAS 09375-6951 could either contain an AGN which is obscured at
optical wavelengths or a complex of supernovae remnants (e.g. Arp220;
Smith et al. 1998b).  As discussed in Section 2.2.2, Kewley
et al. (2000) found that the sources they believed to be complexes of
SNR tended to have compact cores with luminosities of $<10^{4}{\rm
L}_{\sun}$. The $0.05''$ compact core in IRAS 09375- 6951 
has a luminosity which would place it among the AGN rather than the starburst galaxies.

\item {\bf IRAS 13135-2801 (NGC 5051)} (HII; $q$=2.91) is an SA(rs)b galaxy with a
nuclear ring \citep{Buta95}. We detect a compact core in both
polarizations at the level of 2.2\,mJy ($10^{3.75}$ L$\sun$). This is
the only object with a compact radio core which exhibits a {\it
deficit} in radio flux, emitting 17\% {\it less} radio flux than would
be predicted from the FIR luminosity. The optical line ratios exhibited by this galaxy place it in the starburst of HII regime. The low luminosity of the
radio core combined with the optical spectroscopic classification
means that it is highly possible that we have detected a complex of
radio supernovae in this object rather than an AGN.

\item {\bf IRAS 13097-1531 (NGC 1510)} (HII; $q$=2.75) is an S0 (or
peculiar spiral) and is known to contain an OH megamaser
\citep{Martin89}. It has been
classified from optical spectra as a ``composite object'' by \citet{Baan98} (who used a similar classification system to V95) and our own
observations show that it lies close to the AGN/HII region
partition in the V95 classification system. Under the K01 classification system it is classified as a starburst (HII) galaxy. A compact radio core at the level of 2.5\,mJy was detected (Paper I). IRAS 13097-1531 is similar to IRAS 09375-6951
in that the optical classification is ambiguous but the radio
luminosity of the core is $10^{4.14}$ L$\sun$, brighter than would be
expected for the supernova complexes.
\end{itemize}

\end{document}